\documentclass[11pt]{amsart}
\usepackage{amsaddr}
\usepackage{geometry}                
\usepackage{graphicx}
\usepackage{amssymb}
\usepackage{epstopdf}
\usepackage{amssymb}
\usepackage{amsmath}
\usepackage{blkarray} 
\usepackage{latexsym}
\usepackage{amsthm} 
\usepackage{eucal} 
\usepackage{eufrak}
\usepackage{float}
\usepackage{graphicx}
\usepackage[utf8]{inputenc}
\usepackage[main=english, ukrainian]{babel}
\usepackage{xcolor}
\usepackage{mathrsfs}
\usepackage{array}
\usepackage{framed}
\usepackage[all]{xy}
\usepackage{sidecap}
\usepackage{wrapfig}
\usepackage{multicol}
\usepackage{lscape}
\usepackage{MnSymbol}
\usepackage{mathtools}
\usepackage{cite}
\usepackage{appendix}
\usepackage{chngcntr}
\usepackage{etoolbox}
\usepackage{lipsum}
\usepackage{rotating}
\usepackage[thinlines]{easytable}
\usepackage{hyperref}
\usepackage{empheq}
\usepackage{enumitem}
\usepackage{lineno}
\DeclareMathAlphabet{\mathpzc}{OT1}{pzc}{m}{it}

\theoremstyle{plain} 
\newtheorem{thm}{Theorem}[section]

\newtheorem{prop}[thm]{Proposition} 

\theoremstyle{definition} 
\newtheorem{defn}{Definition}

\theoremstyle{remark} 

\newtheorem{rmk}{Remark}

\title[Sign-sensitivity of metabolic networks]{Sign-sensitivity of metabolic networks:\\
which structures determine\\ 
the sign of the responses}
\author{Nicola Vassena}
\address{Freie Universit{\"a}t Berlin}
\email{nicola.vassena@fu-berlin.de}
\date{\today}

\begin{document}

\providecommand{\keywords}[1]
{
  \small	
  \textbf{\textit{Keywords---}} #1
}

\maketitle

\tableofcontents

\begin{abstract}
Perturbations are ubiquitous in metabolism. {A central tool to understand and control their influence on metabolic networks is sensitivity analysis}, which investigates how the network responds to external perturbations. We follow here a structural approach: the analysis is based on the network stoichiometry only and it does not require any quantitative knowledge of the reaction rates. We consider perturbations of reaction rates and metabolite concentrations, at equilibrium, and we investigate the responses in the network.\\
For general metabolic systems, this paper focuses on the sign of the responses, i.e. whether a response is positive, negative or whether its sign depends on the parameters of the system. {In particular, we identify and describe the subnetworks that are the main players in the sign description. These subnetworks are associated to certain kernel vectors of the stoichiometric matrix and are thus independent from the chosen kinetics.}

\smallskip

\end{abstract}

\section{Introduction}
A typical problem in biological networks is to understand how such systems respond to perturbations. Perturbations may be induced both by environmental as well as genetic agents, and the interest is at least twofold. Firstly, it is of interest investigating the \emph{robustness} of a system, that is, the maintenance of certain dynamical properties under the effect of external perturbations. Secondly, and quite on the contrary, it is of interest developing strategies to \emph{influence} and \emph{control} such dynamical properties by precise targeted perturbations, as many medical and pharmacological applications are based on related concepts \cite{Lopez2008, Zapetal21}.\\ 
{Control of living matters is a very delicate task. Firstly, it is very difficult to carry out a perturbation in such fragile context, even more if a perturbation must be targeted to a single network component. An interesting method of metabolism perturbation is studied in the fundamental contribution \cite{Ish07} by Ishii et al, where the authors perform enzyme knock-out experiments on the central glucose metabolism of  \emph{Escherichia Coli}. Via a genetic modification of the cell's DNA, the gene responsible for producing an enzyme catalyzing a certain reaction $j^*$ is removed, so that the rate of the reaction $j^*$ is decreased. The responses in the network, obtained experimentally, showed an intriguing pattern feature: many components of the network did not respond, at all, and many others showed intercorrelated responses. A second more general mathematical difficulty is the lack of precise and reliable quantitative values, as measurements are often very difficult in applications. A possibility to overcome such intrinsic difficulty is the employment of \emph{qualitative approaches} rather than quantitative numerical simulations based on uncertain parameters. This raises the mathematical question, whether the signs of the responses can be understood by the structure of the network, alone. The goal of this paper is to identify the network structures encoding the responses and thus possibly explaining the arising of the patterns. 
More specifically, some of the questions this paper deals with are:
\begin{enumerate}
\item Which reaction $j^*$ must be perturbed to influence the flux of a reaction $j'$? 
\item To reach a positive influence, which should be the sign of the perturbation? 
\item For a fixed perturbation, can the sign of the influence be controlled via a careful choice of the reaction rate parameters?
\end{enumerate}}

{In a dynamical systems context, as it is of interest of the present contribution, the following system of Ordinary Differential Equations (ODE) is typically considered:
\begin{equation} \label{primo}
\dot{x}=f(x):=S\mathbf{r}(x),
\end{equation}
where $x \ge 0$ is the vector of the concentrations of chemicals or metabolites, $S$ is the stoichiometric matrix, and $\mathbf{r}(x)$ is the vector of the reaction rates (\emph{kinetics}). Let us consider a network at a \emph{positive equilibrium} $\bar{x}>0$:
\begin{equation}
0=f(\bar{x}).
\end{equation}
To include external perturbations, a further $\varepsilon$-dependence is added:
\begin{equation} \label{primopert}
0=f(\bar{x}, \varepsilon),
\end{equation}
where the perturbation $\varepsilon \ge 0$ may as well be interpreted as a control term. Assume that there exists a positive equilibrium $\bar{x}(\varepsilon)>0$ for $\varepsilon$ in a neighborhood of 0. The central object of sensitivity analysis are the partial derivatives of the responsive components with respect to $\varepsilon$, at $\varepsilon=0$. The concentrations of metabolites and the reaction fluxes are natural responsive components to be considered in a metabolic network case. The \emph{concentration response} of metabolite $m'$ to an $\varepsilon$-perturbation is defined as
\begin{equation}\label{primomet}
\delta x_{m'}:=\frac{\partial \bar{x}_{m'}(\varepsilon)}{\partial \varepsilon}\bigg|_{\varepsilon=0},
\end{equation}
and the \emph{flux response} of reaction $j'$ to an $\varepsilon$-perturbation is
\begin{equation}\label{primofluxflux}
\Phi_{j'}:=\frac{\partial r_{j'} (\bar{x}(\varepsilon))}{\partial \varepsilon}\bigg|_{\varepsilon=0}.
\end{equation}
}

{Our results focus on metabolic networks as intended applications. Mathematically, equations like \eqref{primo} model also more general chemical reaction networks and even ecological and epidemiological systems. However, the choice of the class of functions $\mathbf{r}$ and of the stoichiometric matrix $S$ may be very different. In the frame of ecology, for example, \cite{Yo88, NA92} studied a sensitivity matrix for \emph{`food webs'} and \emph{`flow networks`}.} In the frame of chemistry, several types of sensitivity analysis have been investigated, following both local and global approaches. We refer to the survey paper \cite{SRTC05} for an overview of these methods and more detailed references. In an ODE context, Shinar and Feinberg investigated a property called \emph{absolute concentration robustness} (ACR) \cite{ShiFei09, ShiFei10, ShiFei11}. In the authors' words \cite{ShiFei11}, ``a model biochemical system has ACR relative to a particular bio-active molecular species if [...] the concentration of that species is the same in all of the positive steady states that the system might admit, regardless of the overall supplies of the various network constituents''. ACR may indicate zero sensitivity of the concentration of a certain species with respect to the other network components. However, the precise mathematical connections between ACR and zero sensitivity are still to be investigated. {For a first few attempts to establish such a bridge, see \cite{PeF20, Capet20}.} Moreover, in \cite{Shietal11} Shinar and co-authors were able to derive quantitative bounds on the entries of the sensitivity matrix for reaction fluxes, in a mass-action kinetics context and for a regular class of networks. To our knowledge, only few contributions further address the \emph{signs} of the sensitivity responses \cite{Sontag14, Gio16, Feliu19}.\\

{With a focus on metabolism and on gene knock-out experiments studied in \cite{Ish07}, Fiedler and Mochizuki \cite{MF15, FM15} started a structural sensitivity analysis of equilibria. This body of work has subsequently been enlarged by further contributions \cite{BF18, VM17, V17, V21}. Knock-out experiments forbid the production of an enzyme, perturbing thus the rate of the corresponding reaction. Consequently the following \emph{reaction perturbation} was considered:
\begin{equation} \label{perteq}
0=f(\bar{x}, \varepsilon)=S \mathbf{r}^\varepsilon(\bar{x}),
\end{equation}
where $\mathbf{r^{\varepsilon}}(x)=(1+\varepsilon e_{j^*})\mathbf{r}(x)$ and $e_{j^*}$ is the $j^*$-th unit vector in $\mathbb{R}^N$. \eqref{perteq} models a targeted perturbation of the single rate of reaction $j^*$. Without specifying the kinetics, only algebraic relations between network components can be addressed. The responses \eqref{primomet} and \eqref{primofluxflux} have been termed \emph{algebraically nonzero}, if they are non identically zero upon differentiation. An algebraically nonzero response of an element $p'$, either a metabolite $m'$ or a reaction $j'$, to a perturbation of $j^*$ has been called \emph{nonzero influence of $j^*$ on $p'$}, and denoted with the notation:
$$j^* \quad  \rightsquigarrow \quad  p'.$$
Moreover, \cite{BF18} proved transitivity of reaction influence:
$$j_1 \rightsquigarrow j_2 \rightsquigarrow j_3 \quad \Rightarrow \quad j_1 \rightsquigarrow j_3.$$}

This paper addresses for the first time in this metabolic context the question on the \emph{sign} of the responses. For a reaction perturbation of a single reaction $j^*$, at equilibrium, we ask
$$\textit{What is the sign of the responses?}$$
{We focus on flux responses $\Phi$, for mathematical reasons: following our approach, treating flux responses is less technical and more intuitive, so that it is preferable as a first exposition to the topic. We have thus omitted a full analysis of the metabolite responses $\delta x$ for concision and clarity of presentation. For an attempt to the sign-analysis of $\delta x$ along the lines of the present paper, see the dissertation \cite{Vas20}}.\\

A central concept in our approach are the \emph{Child Selections}. A Child Selection $\mathbf{J}$ is an injective map associating to each metabolite $m$ a reaction $j$, in which the metabolite $m$ participates as input reactant, see Definition \ref{CSDEF} below. In particular, a Child Selection identifies reshuffled square minors $S^\mathbf{J}$ of the stoichiometric matrix $S$, such that the $i^{th}$ column of $S^\mathbf{J}$ is the $\mathbf{J}(m_i)$ column of $S$. These minors play a crucial role in the structural description of the nonzero response.\\

{Let $r_{jm}$ indicate the partial derivative of the $j^{\text{th}}$ reaction rate $r_j$ with respect to the concentration of the $m^{\text{th}}$ metabolite, i.e., $r_{jm}:=\frac{\partial {r_{{j}}}(x)}{\partial x_m}$. Introduced in \cite{BF18}, the formula for the flux response $(\Phi)^{j^*}_{j'}$ of reaction $j'$ to a reaction perturbation of $j^* \neq j'$ reads
\begin{equation}\label{jiajiajia}
\operatorname{det}SR \cdot (\Phi)^{j^*}_{j'}=\sum_{{j^*}\not\in\mathbf{J}\ni{j'} } (\varphi^\mathbf{J})^{j^*}_{j'},
\end{equation}
where the square matrix $SR$ is the Jacobian matrix of the system \eqref{primo}, $\mathbf{J}$ are Child Selections, and $(\varphi^\mathbf{J})^{j^*}_{j'}$ are multilinear homogeneous monomials in the variables $r_{jm}$ with a coefficient that depends on certain minors of the stoichiometric matrix $S$, identified via $\mathbf{J}$, see \eqref{drjia2} below. Hence, any response can be expressed as a \emph{rational function} and the sign of the flux response $(\Phi)^{j^*}_{j'}$ depends both on the sign of the Jacobian determinant $\operatorname{det}SR$ and on the sign of each \emph{response summand} $(\varphi^\mathbf{J})^{j^*}_{j'}$. Since a structural analysis of the sign of the Jacobian $\operatorname{det}SR$ has already been done in \cite{VGB20}, in the present paper we address the sign of the response summands $(\varphi^\mathbf{J})^{j^*}_{j'}$. With a natural monotonicity assumption on the kinetics, i.e. $r_{jm}>0$, addressing the sign of $(\varphi^\mathbf{J})^{j^*}_{j'}$ as a monomial of the variables $r_{jm}$ means investigating the coefficient of such monomial.}\\

The main tool in the analysis are the \emph{Enlarged Child Selections} (ECS) $\mathbf{J}\cup j^*$, for $j^* \not \in \mathbf{J}$. An ECS naturally identifies an $M\times (M+1)$ matrix $S^{\mathbf{J}\cup j^*}$, where $j^*$ is the column $M+1$ and the first $M$ columns are identical to $S^{\mathbf{J}}$. The two main results of this paper, contained in Section \ref{SRMR}, fix a Child Selection $\mathbf{J}$ and describe the sign of $(\varphi^\mathbf{J})^{j^*}_{j'}$ for any $j' \in \mathbf{J}$. Specifically, Proposition \ref{basic} shows that the only relevant case is when the dimension of the kernel of $S^{\mathbf{J}\cup j^*}$ is exactly one: trivial kernels are excluded by the dimension $M \times (M+1)$ of $S^{\mathbf{J}\cup j^*}$ and kernels of dimension bigger than one indicate zero response summands $(\varphi^\mathbf{J})^{j^*}_{j'}$, for all $j'$. The analysis highlights in particular the important role played by nonzero ECS kernel vectors $0 \neq v \in \mathbb{R}^{M+1}$,
\begin{equation}
S^{\mathbf{J}\cup j^*} v = 0,
\end{equation} 
in the one-dimensional kernel situation, $\operatorname{ker}(S^{\mathbf{J}\cup j^*})=\operatorname{span}\langle v\rangle$. In this case, 
\begin{center}
\textit{The sign pattern of the entries $v_j$ holds the key to the sign pattern of the responses.}
\end{center}
In fact, Theorem \ref{signres1} states that nonzero response summands $(\varphi^\mathbf{J})^{j^*}_{j'} \neq 0$ are characterized by nonzero entries $v_{j'} \neq 0$, and the mutual sign of the entries translates to the mutual sign of the response summands. That is, for $j'_1$ and $j'_2$,
\begin{equation}
\operatorname{sign} ( (\varphi^\mathbf{J})^{j^*}_{j'_1} (\varphi^\mathbf{J})^{j^*}_{j'_2} ) = \operatorname{sign} (v_{j'_1}v_{j'_2}).
\end{equation}
{The determination of the absolute sign of each summand is then addressed in Theorem \ref{signres2}, requiring a bit of technicality for which we refer directly to Section \ref{SRMR}.}\\

As corollaries of the analysis, valuable qualitative considerations can be drawn. Firstly, \ref{twins} shows that if a metabolite $m$ participates in \emph{only} two reactions $j_1$ and $j_2$, then the responses to a perturbation $j_1$ are identical but opposite in sign to the responses of a perturbation of $j_2$. Secondly, \ref{Signtrans} shows with a simple counterexample that no sign-transitivity result holds, that is,
\[ j_1 \mathrel{\mathop{\rightsquigarrow}^{\mathrm{+}}} j_2 \mathrel{\mathop{\rightsquigarrow}^{\mathrm{+}}} j_3 \not \Rightarrow  j_1 \mathrel{\mathop{\rightsquigarrow}^{\mathrm{+}}} j_3,\]
or any other combination of sign. {To our knowledge, it is the first time that this counterintuitive feature of sign-sensitivity is described.} Thirdly, \ref{indnoind} shows that it may be possible, in \eqref{jiajiajia}, that the Jacobian $\operatorname{det}SR$ on the left hand side and the right hand side $\sum_{{j^*}\not\in\mathbf{J}\ni{j'} } (\varphi^\mathbf{J})^{j^*}_{j'}$ share common polynomial factors.\\

The paper is organized as follows: the settings are presented in more detail in Section \ref{metnet} (metabolic networks) and \ref{sensitivity} (sensitivity). The main results are contained in Section \ref{SRMR}.  Section \ref{exsign} presents four examples. {Section \ref{metpertsec} briefly discusses the case of metabolite perturbation.} Section \ref{discussion} concludes with the discussion. All proofs are listed in \ref{Signproof}.\\

\section{Metabolic networks in mathematics}\label{metnet}

A metabolic network $\mathbf{\Gamma}$ is a pair $\{\mathbf{M},\mathbf{E}\}$, where $\mathbf{M}$ is the set of metabolites and $\mathbf{E}$ is the set of reactions. The cardinality of $\mathbf{M}$ is $M$, i.e., $|\mathbf{M}|=M$ and $N$ is the cardinality of $\mathbf{E}$, i.e., $|\mathbf{E}|=N$. In examples, we use capital letters $A,B,...$ for metabolites and numbers $1,2,...$ for reactions. Letters $m \in \mathbf{M}$ and $j \in \mathbf{E}$ refers generically to metabolites and reactions, respectively.\\

A reaction $j$ is an ordered association between two positive linear combinations of metabolites:
\begin{equation} \label{reactionj}
 j: \quad s^{j}_1m_1+...+s^{j}_Mm_M \underset{j}{\longrightarrow} \tilde{s}^{j}_1m_1+...+\tilde{s}^{j}_Mm_M.
\end{equation}
The nonnegative coefficients $s^{j},\tilde{s}^j$ are called \emph{stoichiometric} coefficients. In metabolic networks, these stoichiometric coefficients are integer and mostly 0 or 1. Mathematically, there is no need to impose such restriction, and we can freely consider real $s^{j}_m,\tilde{s}^j_m \in {\mathbb{R}_{\ge 0}}$.\\
If a metabolite $m$ appears at the left hand side of \eqref{reactionj} with nonzero coefficient, then we say that $m$ is an \emph{input} of reaction $j$. Conversely, if $m$ appears on the right hand side with nonzero coefficient, we call $m$ an $\emph{output}$ of reaction $j$. Naturally, metabolic systems are open systems, exchanging chemicals with the outside environment via inflows and outflows. In this context, \emph{inflow reactions} are then reactions with no inputs ($s^j=0$) and \emph{outflow reactions} are reactions with no outputs ($\tilde{s}^j=0$).\\

The $M \times N$ stoichiometric matrix $S$ encodes all the ordered stoichiometric coefficients:
\begin{equation}\label{smatrix}
S_{mj}:=
\begin{cases}
-s^j_m\,\;\quad \text{ for $m$ input of $j$},\\
\tilde{s}^j_{m}\quad\quad \text{ for $m$ output of $j$},\\
0\quad\quad\quad\text{if $m$ does not participate in reaction $j$}.\\
\end{cases}
\end{equation}
Throughout this paper, we always use the notation $S^j$ to refer to the column of the stoichiometric matrix $S$ associated to the reaction $j$. For example, in a network of four metabolites $\{A,B,C,D\}$, an outflow reaction from metabolite $A$ is represented as the $j^{th}$ column of the stoichiometric matrix $S$ as
\begin{equation}
S^j=
\begin{blockarray}{cc}
 & j \\
\begin{block}{c(c)}
  A & -1\\
  B &  0\\
  C & 0\\
  D & 0\\
\end{block}
\end{blockarray}\;.
\end{equation}
Note that stoichiometric columns associated to inflow reactions always have only negative entries. On the contrary, columns associated to inflow reactions have only positive entries. With this construction a fixed order is assigned to each reaction. In particular, we model a reversible reaction
\begin{equation}
j: \quad A+B \underset{j}{\longleftrightarrow} 2C
\end{equation}
simply as two irreversible reactions
\begin{equation}
j_1: \quad A+B \underset{j_1}{\longrightarrow} 2C\quad \text{ and }\quad j_2: \quad 2C \underset{j_2}{\longrightarrow} A+B.
\end{equation}

Let now $x \ge 0$ be the $M$-vector of the concentrations of metabolites. Under the assumption that the reactor is well-mixed, spatially homogeneous and isothermal, the dynamics $x(t)$ of the concentrations satisfy the following system of ODEs:
\begin{equation} \label{ODE}
\dot{x}=f(x)=S\mathbf{r}(x),
\end{equation}
where $S$ is the $M \times N$ stoichiometric matrix \eqref{smatrix} and $\mathbf{r}(x)$ is the $N$-vector of the \emph{reaction rates} (kinetics). We do not require any specific form of such kinetics functions.
We consider the reaction rate of inflow `feed' reactions $j_f$, with no inputs at all, as constant,
\begin{equation}
r_{j_f}(x)\equiv K_{j_f}.
\end{equation}
For any other reaction $j$, $r_j(x)$ is a {nonnegative} and monotonically increasing $C^1$ function that depends only on the concentrations of those metabolites that are input to $j$:
\begin{equation}
\frac{\partial r(x)} {\partial x_m} \equiv 0  \quad \quad \quad \Leftrightarrow \quad \quad \quad s_m^j=0.
\end{equation}
If the metabolite $m$ is an input to the reaction $j$, the notation $r_{jm}$ indicates the positive partial derivative
\begin{equation}
r_{jm}:= \frac{\partial r_j(x)}{\partial x_m} > 0.
\end{equation}
The monotonicity restriction is indeed satisfied by most, but not all, chemical reaction schemes. Without any constraints on the sign of $r_{jm}$, we will not be able to predict the sign of the responses, of course.\\

Many and fundamental questions have arisen in literature connected to the existence, uniqueness, and stability of equilibria solutions $\bar{x}$ of \eqref{ODE}
\begin{equation} \label{ODEeq}
0=f(\bar{x})=S\mathbf{r}(\bar{x}),
\end{equation}
for which we refer to the comprehensive book of Martin Feinberg \cite{Fei19}. In this paper, we assume a priori the existence of a {positive} equilibrium $\bar{x}>0$ of \eqref{ODEeq}, but we require neither its uniqueness nor its stability. {In our setting, the existence assumption characterizes the stoichiometric matrix $S$ possessing a positive \emph{flux} kernel vector $\mathbf{r}$. In particular,  \eqref{ODEeq} imposes linear constraints on the reaction rate functions.} Note that these constraints do not necessarily fix the precise value of an equilibrium $\bar{x}$, and can be considered posed a priori, so that the existence of an equilibrium is really an assumption on the reaction rates $\mathbf{r}(x)$, only. However, the analysis in this paper is entirely based on the derivatives $r_{jm}$ of the reaction rates, and we must issue a warning here: depending on the parametric richness of the kinetics, {i.e., which class of nonlinearities we consider}, the derivatives $r_{jm}$ may or may not be considered {parametrically} independent from each other and from the linear constraints \eqref{ODEeq}. For example, for polynomial mass action kinetics, the value of $r_j(x)$ and $r_{jm}(x)$ are related, a priori, at any value $x$, and for any $j$ and $m$. In contrast, Michaelis-Menten kinetics possesses enough parametric freedom to consider the partial derivatives $r_{jm}$ as positive parameters, independent from \eqref{ODEeq}. This argument is computed explicitly in \cite{VGB20} and we omit it here. We will address again this topic in the discussion section \ref{discussion}.

\section{Sensitivity setting}\label{sensitivity}

In this paper we focus on perturbations of the reaction rates {at a positive equilibrium $\bar{x}>0$}. We consider the following reaction-perturbed equation:
\begin{equation}\label{perteq2}
0=f(\bar{x},\varepsilon)=S \mathbf{r}^\varepsilon(\bar{x}),
\end{equation}
{where $\mathbf{r^{\varepsilon}}(x)=(1+\varepsilon e_{j^*})\mathbf{r}(x)$ and $e_{j^*}$ is the $j^*$-th unit vector in $\mathbb{R}^N$. The perturbation thus concerns the rate of the single reaction $j^*$.} 
At $\varepsilon=0$, the Jacobian of the system is the matrix $\frac{\partial f}{\partial x}=SR$, where $S$ is again the stoichiometric matrix and $R$ is the $N \times M$ matrix of the partial derivatives $r_{jm}$, 
\begin{equation} \label{Req}
R_{jm}:=\frac{\partial}{\partial x_m} r_j(x) =
\begin{cases}
r_{jm}\quad \text{if}\quad\frac{\partial r_j(x)}{\partial x_m}\neq 0\\
0\quad\quad\text{otherwise}
\end{cases}.
\end{equation}

Under the nondegeneracy assumption, assumed throughout, 
\begin{equation} \label{nondeg}
\operatorname{det}SR\neq 0,
\end{equation}
 the Implicit Function Theorem (IFT) guarantees the existence of a family of equilibria solutions $\bar{x}(\varepsilon)$, 
\begin{equation} \label{perteqeq}
S\mathbf{r}^\varepsilon (\bar{x}(\varepsilon))= 0,
\end{equation}
for $\varepsilon$ in a neighborhood of zero. We refer to Section \ref{discussion} for a discussion about the standing assumption \eqref{nondeg}.\\ 

By differentiation of \eqref{perteqeq}, with respect to $\varepsilon$, we obtain
\begin{equation}\label{perteqdiff}
 0=S\big(e_{j^*}+R \frac {\partial \bar{x}(\varepsilon)}{\partial \varepsilon}\big).
\end{equation}
The \emph{concentration response} of metabolite $m'$ to a perturbation of reaction $j^*$ is defined as
\begin{equation}\label{fluxmetresp}
\delta x^{j^*}_{m'}:=\frac{\partial \bar{x}_{m'}(\varepsilon)}{\partial \varepsilon}\bigg|_{\varepsilon=0}=-[(SR)^{-1}S^{j^*}]_{m'},
\end{equation}
and the \emph{flux response} of reaction $j'$ to a perturbation of reaction $j^*$ as
\begin{equation}\label{fluxfluxresp}
\Phi^{j^*}_{j'}:=\frac{\partial r_j (\bar{x}(\varepsilon))}{\partial \varepsilon}\bigg|_{\varepsilon=0}=\delta_{j^*j'}+(R \delta x^{j^*})_{j'}=\delta_{j^*j'} - [R(SR)^{-1}S^{j^*}]_{j'},
\end{equation}
where $\delta_{j^*j'}$ is the Kronecker-delta. {In vector notation, \eqref{perteqdiff} is then the \emph{flux balance}
\begin{equation}
0= S \Phi^{j^*},
\end{equation}
indicating that the flux responses are kernel vectors of the stoichiometric matrix $S$.} In \cite{BF18}, expressions \eqref{fluxmetresp} and \eqref{fluxfluxresp} have been analyzed in terms of \emph{Child Selections}, whose definition we recall here:
\begin{defn}[Child Selections] \label{CSDEF}
A \emph{Child Selection} is an injective map $\mathbf{J}: \textbf{M} \longrightarrow \textbf{E}$, which associates to every metabolite $m \in \textbf{M}$ a reaction $j \in \textbf{E}$ such that $m$ is an input metabolite of reaction $j$.\\
\end{defn}

{For simplicity of notation, $j\in \mathbf{J}$ indicates that the reaction $j$ is in the image of the map $\mathbf{J}(\mathbf{M})$}. The Jacobian determinant of the system, $\operatorname{det}SR$, can be expanded along Child Selections (Proposition 2.1 of \cite{VGB20}) as:
\begin{equation}\label{detexp}
\operatorname{det}SR=\sum_\mathbf{J} \operatorname{det}S^\mathbf{J} \cdot \prod_{m \in \mathbf{M}} r_{\mathbf{J}(m)m},
\end{equation}
where $S^\mathbf{J}$ indicates the {$M\times M$} matrix whose $m^{th}$ column is the $\mathbf{J}(m)^{th}$ column of $S$. {The sum runs over \emph{all} possible Child Selections}. Note that the existence of a Child Selection $\mathbf{J}$ such that 
\begin{equation*}
\operatorname{det}(S^\mathbf{J}) \neq 0,
\end{equation*}
characterizes $\operatorname{det}SR \not \equiv 0$, as a function of the variables $r_{jm}$. {In this case, of course, $\operatorname{det}SR$ may still be zero for certain values of $r_{jm}$. On the other hand, $\operatorname{det}SR \neq 0$ necessarily implies the existence of a Child Selection $\mathbf{J}$ with $\operatorname{det}(S^\mathbf{J}) \neq 0$}. In addition, possible zeros of $\operatorname{det}SR$ may hint at saddle node bifurcations of equilibria leading to multistability regions, so that it is of interest the sign of each coefficient $\operatorname{det}S^\mathbf{J}$, {in order to detect or exclude a priori the possibility of having a zero of $\operatorname{det}SR$.} This is the content of the next definition.

{\begin{defn}[Child Selection behavior] \label{CSclass}
Let $\mathbf{J}$ be a Child Selection. We define the \emph{behavior coefficient} $\beta(\mathbf{J})$ as
\begin{equation}
\beta(\mathbf{J}):=\operatorname{sign}(\operatorname{det}S^\mathbf{J}).
\end{equation}
Moreover, we say that $\mathbf{J}$ is \emph{good} if $\beta(\mathbf{J})= (-1)^M$; $\mathbf{J}$ is \emph{bad} if $\beta(\mathbf{J})=(-1)^{M-1}$; $\mathbf{J}$ \emph{zero-behaves} if $\beta(\mathbf{J})=0$.\\
\end{defn}}

{Consider a metabolic network admitting a positive stable equilibrium for all reaction rates. The Jacobian of such an equilibrium has either eigenvalues with negative real part or pairs of purely imaginary eigenvalues that are complex conjugated. The sign of a nonsingular Jacobian, then, is always
$$\operatorname{sign}\operatorname{det}SR= (-1)^M,$$
which is implied, via \eqref{detexp}, if all nonzero-behaving Child Selections are \emph{good}. This is why the Child Selections with this sign have been named `good'. Conversely, a loss of stability of an equilibrium, for example via a saddle node bifurcation, necessarily requires at least one \emph{bad} Child Selection.}
Note that a Child Selection naturally identifies a subnetwork {$\{\mathbf{M}, \mathbf{J}(\mathbf{M})\}$ constituted by all the metabolites $m\in\mathbf{M}$ and the reactions $j \in \mathbf{J}$.}  For a given $\mathbf{J}$, \cite{VGB20} contains a network characterization of the behavior coefficient $\beta(\mathbf{J})$ based on certain cycles in the network. {As a consequence of the analysis, for example, certain classes of Child Selections, combinatorially simple, are shown to be good; e.g. acyclic Child Selections are always good.}\\

In \cite{BF18}, a formula for the flux response $(\Phi)^{j^*}_{j'}$ of reaction $j'$ to a reaction perturbation of $j^* \neq j'$ was derived, in Child Selection terms:
\begin{equation} \label{drjia3}
\operatorname{det}SR \cdot (\Phi)^{j^*}_{j'}=\sum_{{j^*}\not\in\mathbf{J}\ni{j'} } (\varphi^\mathbf{J})^{j^*}_{j'},
\end{equation}
where
\begin{equation} \label{drjia2}
(\varphi^\mathbf{J})^{j^*}_{j'} :=-\operatorname{det}S^{\mathbf{J}\setminus j' \cup j^*} \prod_{m\in M} r_{\mathbf{J}(m)m}.
\end{equation}
In \eqref{drjia3} the sum runs over {all} the Child Selections $\mathbf{J}$ selecting the responsive reaction $j'$ as image of one metabolite, but not selecting the perturbed reaction $j^*$. In \eqref{drjia2}, $S^{\mathbf{J}\setminus j' \cup j^*}$ indicates the $M \times M$ matrix obtained from $S^\mathbf{J}$ by replacing the stoichiometric column $S^{j'}$, associated to reaction $j'$, with the column $S^{j^*}$, associated to reaction $j^*$.\\

{Formula \eqref{drjia3} implies that the sign of the response $(\Phi)^{j^*}_{j'}$ depends on the sign of each respond summand $(\phi^\mathbf{J})^{j^*}_{j'}$ \emph{and} on the sign of the Jacobian determinant $\operatorname{det}SR$. Both $(\phi^\mathbf{J})^{j^*}_{j'}$ and $\operatorname{det}SR$ can be abstractly seen as multilinear homogenous polynomials in the positive variables $r_{jm}$: the response $(\Phi)^{j^*}_{j'}$ is thus a rational function of such variables. Consequently, the last definition is concerned with the dependence of the sign of $(\Phi)^{j^*}_{j'}$ on the values of the positive derivatives $r_{jm}$.
\begin{defn}\label{detsigndef}
A response $(\Phi)^{j^*}_{j'}$ is called of \emph{determinate sign} if its sign does not depend on the values of the positive derivatives $r_{jm}$. On the contrary, an \emph{indeterminate sign} response occurs when the sign does depend on the values of $r_{jm}$.
\end{defn}
\begin{rmk}
Definition \ref{detsigndef} is purely algebraic and it is based only on the network structure. In particular, it is independent from the chosen kinetics and the value of the equilibrium $\bar{x}$. We discuss further in the discussion Section \ref{discussion} how this definition plays a role in the interpretation of the results.
\end{rmk}
}

\section{Main results} \label{SRMR}

We recall the \emph{Extended Child Selections} $\mathbf{J}\cup{j^*}$ for $j^* \not \in \mathbf{J}$, and the notation $S^{\mathbf{J}\cup{j^*}}$, indicating the $M\times (M+1)$ matrix possessing $j^*$ as the $(M+1)^{th}$ column and the first $M$ columns identical to $S^{\mathbf{J}}$. We are now ready to present the main results.\\

The first proposition provides a necessary condition for any nonzero response summand $(\varphi^\mathbf{J})^{j^*}_{j'}\neq 0$.
\begin{prop}\label{basic}
For a response summand $(\varphi^\mathbf{J})^{j^*}_{j'}$, it holds:
\begin{equation}
(\varphi^\mathbf{J})^{j^*}_{j'}=0\text{ for all $j'$} \quad \Leftrightarrow \quad \operatorname{dim}(\operatorname{ker}(S^{\mathbf{J}\cup{j^*}}))>1.
\end{equation}
\end{prop}

We now state the first main theorem, on the relative sign of the responses.

\begin{thm}[Relative sign of responses] \label{signres1}

Suppose $\operatorname{dim}(\operatorname{ker}(S^{\mathbf{J}\cup{j^*}}))$=1, and let $\operatorname{ker}S^{\mathbf{J}\cup{j^*}}=\operatorname{span}\langle v \rangle$. Then,
\begin{enumerate}[itemsep=0pt]
\item The response summand of reaction $j'$ is nonzero if and only if the $j'$-th entry of $v$ is nonzero, that is
\begin{equation}
(\varphi^\mathbf{J})^{j^*}_{j'}\neq0 \quad \Leftrightarrow \quad v_{j'}\neq0.
\end{equation}
\item The mutual sign of the response summands of reactions $j'_1$ and $j'_2$ is given by the mutual sign of the $j'_1$-th and $j'_2$-th entries of $v$, that is
\begin{equation}
\operatorname{sign}(\varphi^\mathbf{J})^{j^*}_{j'_1}\operatorname{sign}(\varphi^\mathbf{J})^{j^*}_{j'_2}=\operatorname{sign}(v_{j'_1}v_{j'_2}).
\end{equation}
\end{enumerate}
\end{thm}

To proceed towards the second result of the chapter, on the specific sign of each response, we recall some linear algebra concepts, first, \cite{HJ13}. Let $\mathpzc{A}$ be any $M\times M$ matrix with a one-dimensional kernel. {Straightforwardly, $\mathpzc{A}^T$ has one-dimensional kernel too.} The \emph{cofactor matrix} $\mathpzc{C(A)}$ of $\mathpzc{A}$ is the matrix whose entries $\mathpzc{C(A)}_{mj}$ are given by
\begin{equation}
\mathpzc{C(A)}_{mj}=(-1)^{m+j}\operatorname{det}\mathpzc{A}^{\vee j}_{\,\vee m},
\end{equation}
where $\mathpzc{A}^{\vee j}_{\,\vee m}$ indicates the $(M-1) \times (M-1)$ minor of $\mathpzc{A}$, obtained by removing row $m$ and column $j$. The \emph{adjugate matrix} of $\mathpzc{A}$, $\operatorname{Ad}(\mathpzc{A})$, is then defined as the transpose of the cofactor matrix $\mathpzc{C(A)}$ of A. That is,
\begin{equation}
\operatorname{Ad}(\mathpzc{A})=\mathpzc{C(A)}^T.
\end{equation}
Moreover, we have the relation 
\begin{equation} \label{relation}
\mathpzc{A} \operatorname{Ad}(\mathpzc{A})=\operatorname{Ad}(\mathpzc{A}) \mathpzc{A}=\operatorname{det}\mathpzc{A} \; {\operatorname{Id}_M}=0,
\end{equation}
{where $\operatorname{Id}_M$ is the $M\times M$ identity matrix.} 
Let us fix a kernel vector $v$, which spans $\operatorname{ker}\mathpzc{A}$. Equalities \eqref{relation} imply that {there exists a kernel vector $\kappa=\kappa(v)$ of $\mathpzc{A}^T$, $\operatorname{ker}\mathpzc{A}^T=\operatorname{span}\langle\kappa\rangle$, such that}
\begin{equation}
\operatorname{Ad}(\mathpzc{A}) = \; \,v \cdot \kappa^T\,.
\end{equation}
In particular, any entry $\operatorname{Ad}(\mathpzc{A})_{mj}$ of the adjugate matrix can be expressed as:
\begin{equation}
\operatorname{Ad}(\mathpzc{A})_{mj}=(-1)^{m+j}\operatorname{det}\mathpzc{A}^{\vee m}_{\,\vee j}=v_m \, \kappa_j\,.
\end{equation}
We are now ready to state the second main Theorem \ref{signres2}. 

\begin{thm}[Absolute sign of responses] \label{signres2}
As in Theorem \ref{signres1}, let us suppose $\operatorname{dim}(\operatorname{ker}(S^{\mathbf{J}\cup{j^*}}))$=1, and let $\operatorname{ker} S^{\mathbf{J}\cup{j^*}}=\operatorname{span}\langle v \rangle$. There are two cases:
\begin{enumerate}[itemsep=0pt]
\item If the Child Selection $\mathbf{J}$ does not zero-behave, then the $j^*$-th entry of $v$ is nonzero, i.e. $v_{j^*}\neq 0$, and
\begin{equation} \label{drjia}
\operatorname{sign}(\varphi^\mathbf{J})^{j^*}_{j'}=\beta(\mathbf{J})\operatorname{sign}(v_{j^*}v_{j'}).
\end{equation}
\item  If the Child Selection $\mathbf{J}$ zero-behaves, then $v_{j^*}=0$. In particular, consider $\tilde{v} \in \mathbb{R}^M$ such that $\operatorname{ker}S^\mathbf{J}=\operatorname{span}\langle \tilde{v}\rangle$ and $\tilde{v}_j = v_j$, for any $j=1,...,M$. For the unique {kernel vector $\kappa$ of $(S^\mathbf{J})^T$} such that $\operatorname{Ad}(S^\mathbf{J}) = \; \tilde{v} \cdot \kappa^T$, we have
\begin{equation}\label{rev2}
\operatorname{sign}(\varphi^\mathbf{J})^{j^*}_{j'}=-\operatorname{sign}( v_{j'}\langle\kappa,S^{j^*}\rangle).
\end{equation}
\end{enumerate}
\end{thm}

\begin{rmk}[self-influence]
In analogy to \eqref{drjia3}, \cite{BF18} provided also a formula for the case of self-influence $j'=j^*$. The formula reads
\begin{equation} \label{Bsfluxfluxself}
\operatorname{det}SR\;(\Phi)^{j^*}_{j^*}=\sum_{\mathbf{J}\not\ni j^*} (\tilde{\varphi}^\mathbf{J})^{j^*}_{j^*}
\end{equation}
where the response summands $(\tilde{\varphi}^\mathbf{J})^{j^*}_{j^*}$ are
\begin{equation}
(\tilde{\varphi}^\mathbf{J})^{j^*}_{j^*}=\operatorname{det}S^\mathbf{J}\prod_{m\in \mathbf{M}} r_{\mathbf{J}(m)m}.
\end{equation}
From \eqref{Bsfluxfluxself}, it directly follows that Theorems \ref{signres1} and \ref{signres2} hold also for the case of self-influence $j'=j^*$.  In particular, for this special case, we obtain that
\begin{equation}
\operatorname{sign}(\tilde{\varphi}^\mathbf{J})^{j^*}_{j^*}=\beta(\mathbf{J}).
\end{equation}
\end{rmk}

\subsection{Twin sisters have opposite influence}\label{twins}

As already noted in \cite{BF18}, \eqref{drjia3} implies that if a metabolite $m^*$ participates \emph{only} in one reaction $j^*$, then any flux-response $(\Phi)^{j^*}_{j'}$ to a reaction perturbation of $j^*$ is a priori zero, for any reaction $j'$: the formula requires in particular the existence of a Child Selection $\mathbf{J}$ \emph{not} selecting $j^*$. This interesting feature has been named
\begin{center}
\emph{Single children have no influence.}
\end{center}

Here we add another take-home feature, in this flavor. Note that any ECS $\mathbf{J}\cup j^*$ contains then \emph{at least two} outgoing reactions from a metabolite $m^*$, one of which is $j^*$. We call the reaction $j^*_s=\mathbf{J}(m^*)$ {a} \emph{sister} of $j^*$. Let now $\mathbf{J}$ and $\mathbf{J}_s$ be two Child Selections such that $\mathbf{J}(m)=\mathbf{J}_s(m)$ for any $m \neq m^*$, $\mathbf{J}(m^*)=j_s^*$ and $\mathbf{J}_s(m^*)=j^*$. The matrix $S^{\mathbf{J}_s \setminus j' \cup j^*_s}$ has opposite determinant to the matrix $S^{\mathbf{J} \setminus j' \cup j^*}$. The two matrices are indeed obtainable one from the other, via a single interchange of the columns $S^{j^*}$ and $S^{j^*_s}$. The change of sign in the determinant is a well-known property of a multilinear alternating form. This implies, a priori, that
\begin{equation} \label{lippps}
\operatorname{sign}(\varphi^\mathbf{J})^{j^*}_{j'}=-\operatorname{sign}(\varphi^{\mathbf{J}_s})^{j^*_s}_{j'}.
\end{equation}
Let us now further assume that {any Child Selection $\mathbf{J}$ maps the metabolite $m^*$ either to a reaction $j^*$ or a reaction $j^*_s$, only. That is, }any Child Selection $\mathbf{J}$ is such that $\mathbf{J} (m^*)=j^*$ or $\mathbf{J} (m^*)=j^*_s$. In this case, the statement \eqref{lippps} can be strengthened to the following proposition.

\begin{prop}[Twin sisters have opposite influence] \label{TwinSisters}
Suppose that any Child Selection $\mathbf{J}$ maps the metabolite $m^*$ either to $j^*$ or to $j^*_s$, only. Then
\begin{equation}
r_{j^*m^*}(\Phi)^{j^*}_{j'}=-r_{j^*_sm^*}(\Phi)^{j^*_s}_{j'}, \quad \text{for any $j'$}.
\end{equation}
In particular,
\begin{equation}
\operatorname{sign}(\Phi)^{j^*}_{j'}=-\operatorname{sign}(\Phi)^{j^*_s}_{j'}, \quad \text{for any $j'$}.
\end{equation}
\end{prop}

\section{Examples} \label{exsign}

{In this section we present four examples to illustrate our results and some consequences of our analysis. To help visual intuition, we provide also a graphical representation, where we consider the metabolites as vertices and the reactions as directed hyperarrows, inheriting the orientation from the reaction orientation \eqref{reactionj}. It is one common representation, extensively used in chemistry, biology, and mathematics.}

\subsection{Failure of sign-transitivity of influence} \label{Signtrans} \textcolor{white}{di}\\

Let $j_1, j_2, j_3$ be three distinct reactions. The question of sign-transitivity asks whether the sign of the influence $j_1 \rightsquigarrow j_3$ follows from the signs of the influences $j_1 \rightsquigarrow j_2$, and $j_2\rightsquigarrow j_3$. In symbols,
\[ j_1 \mathrel{\mathop{\rightsquigarrow}^{\mathrm{+}}} j_2 \mathrel{\mathop{\rightsquigarrow}^{\mathrm{+}}} j_3 \xRightarrow{?}  j_1 \mathrel{\mathop{\rightsquigarrow}^{\mathrm{+}}} j_3,\]
or any other combinations of signs. The following toy-model provides a first clear counterexample to sign-transitivity, showing that sign-transitivity fails.
\begin{equation}\label{monotoy}
\includegraphics[scale=0.28]{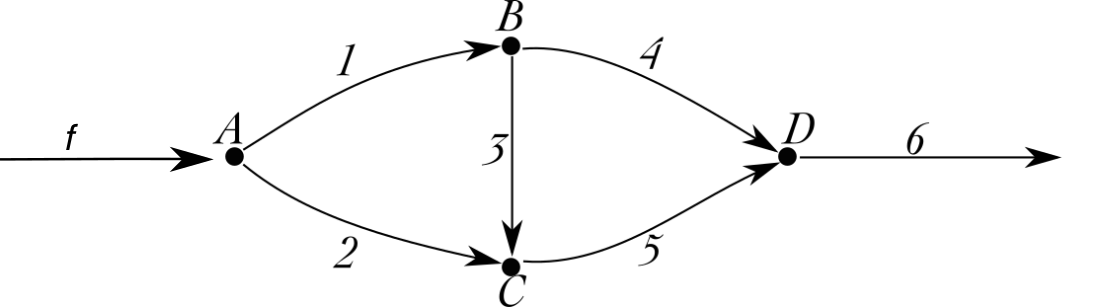}
\end{equation}
\[
S=
\begin{blockarray}{cccccccc}
& f & 1 & 2 & 3 & 4& 5 &6\\
\begin{block}{c[ccccccc]}
  A & 1 & -1 & -1 & 0 & 0 & 0 & 0\\
  B & 0 & 1 & 0 &  -1 & -1 & 0 & 0\\
  C & 0 & 0 &1 & 1 & 0 & -1 & 0\\
  D & 0 & 0 & 0 & 0 & 1& 1 & -1\\
\end{block}
\end{blockarray}\;.
\]
The network \eqref{monotoy} is \emph{monomolecular}, that is, it is composed only by monomolecular reactions $j$ of the form 
\begin{equation} \label{monomoleq}
j: \quad m_1 \underset{j}{\longrightarrow} m_2,
\end{equation}
where a single metabolite $m_1$ reacts to a single metabolite $m_2$. As clarified in \cite{VGB20}, one of the features of monomolecular networks is that the nonsingular Jacobian determinant is always of the `good' sign
\begin{equation}
\operatorname{sign} (\operatorname{det}SR)= (-1)^M,
\end{equation}
and thus for Example \eqref{monotoy}, $\operatorname{sign}(\operatorname{det}SR)= (-1)^4=1$.\\

The positive vector $\mathbf{r}=(3r,2r,r,r,r,2r,3r)^T$, $r \in \mathbb{R}_{> 0}$, is one kernel vector of the stoichiometric matrix $S$, hence the associated dynamical system admits a positive equilibrium $\bar{x}$ and we can perform our sensitivity analysis. We show in detail the following counterexample to sign-transitivity: 
\begin{equation}\label{tasktrans}
 1 \mathrel{\mathop{\rightsquigarrow}^{\mathrm{+}}} 3 \mathrel{\mathop{\rightsquigarrow}^{\mathrm{+}}} 5\text{ but }1 \mathrel{\mathop{\rightsquigarrow}^{\mathrm{-}}} 5.
 \end{equation}

Consider a perturbation of reaction $1$. According to our Theorems, we have first to find Child Selections $\mathbf{J}$ such that $1 \not \in \mathbf{J}$. There are two of such Child Selections, only, depending on {$\mathbf{J}(B)$}:
\begin{equation}
\begin{cases}
\mathbf{J}_3:= \{\mathbf{J}_3(A)=2; \mathbf{J}_3(B)=3; \mathbf{J}_3(C)=5; \mathbf{J}_3(D)=6\}\\
\mathbf{J}_4:= \{\mathbf{J}_4(A)=2; \mathbf{J}_4(B)=4; \mathbf{J}_4(C)=5; \mathbf{J}_4(D)=6\}\\
\end{cases} 
\end{equation}
For both Child Selections it holds:
\begin{equation}
\operatorname{det}S^{\mathbf{J}_i}=1, \quad \quad \quad i=3,4
\end{equation}
Automatically, then, for both Extended Child Selections $\mathbf{J}_3 \cup 1$ and $\mathbf{J}_4 \cup 1$, we have that 
\begin{equation}
\operatorname{dim} \operatorname{ker}S^{\mathbf{J}_i\cup 1}=1, \quad \quad \quad i=3,4,
\end{equation}
and thus Theorem \ref{signres2}, part 1, applies. Let 
$$\operatorname{ker}S^{\mathbf{J}_i\cup 1}= \operatorname{span}\langle v^{\mathbf{J}_i} \rangle, \quad \quad \quad i=3,4.$$
We have that 

\begin{equation}
v^{\mathbf{J}_3}=
\begin{blockarray}{cc}
\begin{block}{c[c]}
  1 &  w\\
  2 &  -w\\
  3 &  w\\
  4 &  0\\
  5 & 0\\
  6 & 0\\ 
\end{block}
\end{blockarray}; \quad \quad \quad
v^{\mathbf{J}_4}=
\begin{blockarray}{cc}
\begin{block}{c[c]}
  1 &  w\\
  2 &  -w\\
  3 &  0\\
  4 &  w\\
  5 & -w\\
  6 & 0\\ 
\end{block}
\end{blockarray},  \quad \quad \quad \text{for } w \in \mathbb{R}.
\end{equation}

Since $v^{\mathbf{J}_3}_1v^{\mathbf{J}_3}_3 > 0$ and $v^{\mathbf{J}_4}_3=0$, we obtain that the response of reaction 3 to a perturbation of reaction 1 is positive:
$$(\Phi)^1_3 = (\varphi^{\mathbf{J}_3})^1_3 + (\varphi^{\mathbf{J}_4})^1_3 = (\varphi^{\mathbf{J}_3})^1_3 > 0,\quad \text{ that is } \quad 1 \mathrel{\mathop{\rightsquigarrow}^{\mathrm{+}}} 3. $$
On the contrary, since $v^{\mathbf{J}_3}_5 = 0 $  and $v^{\mathbf{J}_4}_1 v^{\mathbf{J}_4}_5< 0$, we obtain that the response of reaction {5 to a perturbation of reaction 1 is negative:}
$$(\Phi)^1_5 = (\varphi^{\mathbf{J}_3})^1_5 + (\varphi^{\mathbf{J}_4})^1_5 = (\varphi^{\mathbf{J}_4})^1_5 < 0,\quad \text{ that is } \quad 1 \mathrel{\mathop{\rightsquigarrow}^{\mathrm{-}}} 5.$$

Now consider a perturbation of reaction $3$. To check that $(\Phi)^3_5 > 0$, consider all Child Selections $\mathbf{J}$ such that $3 \not \in \mathbf{J}$. Again, we have only two of those, depending on the image of the metabolite $A$.
\begin{equation}
\begin{cases}
\mathbf{J}_1:= \{\mathbf{J}_1(A)=1; \mathbf{J}_1(B)=4; \mathbf{J}_1(C)=5; \mathbf{J}_1(D)=6\}\\
\mathbf{J}_2:= \{\mathbf{J}_2(A)=2; \mathbf{J}_2(B)=4; \mathbf{J}_2(C)=5; \mathbf{J}_2(D)=6\}\\
\end{cases} 
\end{equation}
In complete analogy as above, for both Child Selections it holds:
\begin{equation}
\operatorname{det}S^{\mathbf{J}_i}=1, \quad \quad \quad i=1,2
\end{equation}
Again, for both Extended Child Selections $\mathbf{J}_1 \cup 3$ and $\mathbf{J}_2 \cup 3$, we have that 
\begin{equation}
\operatorname{dim} \operatorname{ker}S^{\mathbf{J}_i\cup 3}=1, \quad \quad \quad i=1,2,
\end{equation}
and again Theorem \ref{signres2}, part 1, applies. Let 
$$\operatorname{ker}S^{\mathbf{J}_i\cup 3}= \operatorname{span}\langle v^{\mathbf{J}_i} \rangle, \quad \quad \quad i=1,2.$$
Importantly note that for different Child Selections $\mathbf{J}_1, \mathbf{J}_2$, we have 
\begin{equation}\label{smk}
v^{\mathbf{J}_1} = v^{\mathbf{J}_2},\quad \quad \text{ and } \quad \quad
v^{\mathbf{J}_i}=
\begin{blockarray}{cc}
\begin{block}{c[c]}
  1 &  0\\
  2 &  0\\
  3 &  w\\
  4 &  -w\\
  5 & w\\
  6 & 0\\ 
\end{block}
\end{blockarray}; \quad \quad \quad \text{for } w \in \mathbb{R}, \quad i=1,2.
\end{equation}
Since both $v^{\mathbf{J}_1}_3v^{\mathbf{J}_1}_5 > 0$ and $v^{\mathbf{J}_2}_3v^{\mathbf{J}_2}_5 > 0$, we obtain that the response of reaction 5 to a perturbation of reaction 3 is positive:
$$(\Phi)^3_5 = (\varphi^{\mathbf{J}_1})^3_5 + (\varphi^{\mathbf{J}_2})^3_5> 0,\quad \text{ that is } \quad 3 \mathrel{\mathop{\rightsquigarrow}^{\mathrm{+}}} 5. $$
In conclusion, we have showed \eqref{tasktrans}. Other sign combinations follow in analogy. From the same example:
\[ 2 \mathrel{\mathop{\rightsquigarrow}^{\mathrm{-}}} 1 \mathrel{\mathop{\rightsquigarrow}^{\mathrm{-}}} 5, \quad \quad \quad 2 \mathrel{\mathop{\rightsquigarrow}^{\mathrm{+}}} 5.\]

\[ 1 \mathrel{\mathop{\rightsquigarrow}^{\mathrm{+}}} 4 \mathrel{\mathop{\rightsquigarrow}^{\mathrm{-}}} 5, \quad \quad \quad 1 \mathrel{\mathop{\rightsquigarrow}^{\mathrm{-}}} 5,\]
\[ 1 \mathrel{\mathop{\rightsquigarrow}^{\mathrm{+}}} 4 \mathrel{\mathop{\rightsquigarrow}^{\mathrm{-}}} 3, \quad \quad \quad 1 \mathrel{\mathop{\rightsquigarrow}^{\mathrm{+}}} 3.\]

\[ 2 \mathrel{\mathop{\rightsquigarrow}^{\mathrm{-}}} 1 \mathrel{\mathop{\rightsquigarrow}^{\mathrm{+}}} 3, \quad \quad \quad 2 \mathrel{\mathop{\rightsquigarrow}^{\mathrm{-}}} 3.,\]
\[ 2 \mathrel{\mathop{\rightsquigarrow}^{\mathrm{-}}} 3 \mathrel{\mathop{\rightsquigarrow}^{\mathrm{+}}} 5, \quad \quad \quad 2 \mathrel{\mathop{\rightsquigarrow}^{\mathrm{+}}} 5.\]

All remaining possible cases, including indeterminate sign responses, are easily constructible with analogous new examples, and we omit them here.

\subsection{The responses for a zero-behaving Child Selection} \textcolor{white}{di}\\

We exemplify Theorem \ref{signres2}, part 2, by considering a zero-behaving Child Selection $\mathbf{J}$ {of a non-specified network $\mathbf{\Gamma}$ of} seven metabolites: $m^*$, $A,$ $B,$ $C$, $D$, $E$, $F$. 

\begin{equation}
\includegraphics[scale=0.48]{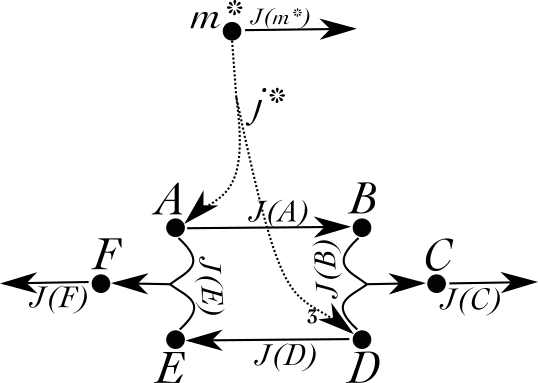}
\end{equation}
\[
S^\mathbf{J}=
\begin{blockarray}{cccccccc}
 & \mathbf{J}(m^*) & \mathbf{J}(A) & \mathbf{J}(B) & \mathbf{J}(C) & \mathbf{J}(D) & \mathbf{J}(E) & \mathbf{J}(F)\\
\begin{block}{c[ccccccc]}
  m^* & -1 & 0 & 0 & 0 & 0 & 0 & 0\\
  A & 0 & -1 & 0 &  0 & 0 & -1 & 0 \\
  B & 0 &1 &  -1 & 0 & 0 & 0 & 0\\
  C & 0 & 0 & 1  &  -1 & 0 & 0 & 0\\
  D & 0 & 0 & -1 & 0 & -1 & 0 & 0\\ 
  E & 0 & 0 & 0  & 0 & 1 & -1 & 0\\
  F & 0 & 0 & 0 & 0 & 0 & 1& -1\\
\end{block}
\end{blockarray}\;,\quad  
S^{j^*}=
\begin{blockarray}{cc}
\begin{block}{c[c]}
  m^* & -1\\
  A & 1\\
  B & 0\\
  C & 0\\
  D & 3\\ 
  E & 0\\
  F & 0\\
\end{block}
\end{blockarray}.   \]
We study the response summands $(\varphi^\mathbf{J})^{j^*}_{j'}$ of reactions $j'=$ $\mathbf{J}(A),$ $\mathbf{J}(B),$ $\mathbf{J}(C),$ $\mathbf{J}(D),$ $\mathbf{J}(E)$, $\mathbf{J}(F)$, to a perturbation of the dashed reaction $j^*$. {They are summands in the responses $\Phi^{j^*}_{j'}$, for $j'=$ $\mathbf{J}(A),$ $\mathbf{J}(B),$ $\mathbf{J}(C),$ $\mathbf{J}(D),$ $\mathbf{J}(E)$, $\mathbf{J}(F)$ respectively. Note that the response summands relative to a Child Selection $\mathbf{J}$ are independent from the full network $\mathbf{\Gamma}$}.\\

The matrix $S^\mathbf{J}$ is singular, and the vector $v=(0, w, w, w, -w, -w, -w)^T$, $w \in \mathbb{R}$, is a kernel vector of $S^\mathbf{J}$. Moreover, 
\begin{equation}
\operatorname{dim}(\operatorname{ker}S^\mathbf{J})=1\quad\quad\text{and thus}\quad\quad\operatorname{ker}S^\mathbf{J}=\operatorname{span}\langle v \rangle.
\end{equation} 
Now, the adjugate matrix $\operatorname{Ad}(S^\mathbf{J})$ is
\begin{equation}
\operatorname{Ad}(S^\mathbf{J})=
\begin{blockarray}{ccccccc}
\begin{block}{[ccccccc]}
   0 & 0 & 0 & 0 & 0 & 0 & 0\\
  0 & 1 & 1 &  0 & -1 & -1 & 0 \\
  0 & 1 & 1 &  0 & -1 & -1 & 0 \\
  0 & 1 & 1 &  0 & -1 & -1 & 0 \\
  0 & -1 & -1 &  0 & 1 & 1 & 0 \\  
  0 & -1 & -1 &  0 & 1 & 1 & 0 \\  
  0 & -1 & -1 &  0 & 1 & 1 & 0 \\
\end{block}
\end{blockarray}\;,
\end{equation}
and the choice $\kappa=(0, \frac{1}{w}, \frac{1}{w}, 0, -\frac{1}{w} -\frac{1}{w}, 0)^T$ of the kernel vector {of $(S^\mathbf{J})^T$} satisfies
\begin{equation}
v \cdot \kappa^T = \operatorname{Ad}(S^\mathbf{J}).
\end{equation} 
For simplicity of computation, we can consider in particular the choice $w=1$, so that $v=(0, 1, 1, 1, -1, -1, -1)^T$ and $\kappa=(0, 1, 1, 0, -1 -1, 0)^T$. Firstly we compute 
\begin{equation}
-\langle\kappa,S^{j^*}\rangle=-1\cdot -2 = +2.
\end{equation}
{According to Theorem \ref{signres2}, the signs of the response summands $(\varphi^\mathbf{J})^{j^*}_{j'}$ for $j'=\mathbf{J}(A)$, $\mathbf{J}(B),$ $\mathbf{J}(C),$ $\mathbf{J}(D),$ $\mathbf{J}(E),$ $\mathbf{J}(F)$, are}:
\begin{equation}
\begin{cases}
\operatorname{sign}(\varphi^\mathbf{J})^{j^*}_{\mathbf{J}(A)}= \operatorname{sign}(-\langle\kappa,S^{j^*}\rangle \, v_{\mathbf{J}(A)}) = \operatorname{sign}(+2 \cdot 1)  > 0. \\
\operatorname{sign}(\varphi^\mathbf{J})^{j^*}_{\mathbf{J}(B)}=  \operatorname{sign}(-\langle\kappa,S^{j^*}\rangle \,v_{\mathbf{J}(B)}) = \operatorname{sign}(+2 \cdot 1)  > 0.  \\
\operatorname{sign}(\varphi^\mathbf{J})^{j^*}_{\mathbf{J}(C)}=  \operatorname{sign}(-\langle\kappa,S^{j^*}\rangle \,v_{\mathbf{J}(C)}) = \operatorname{sign}(+2 \cdot 1)  > 0.  \\
\operatorname{sign}(\varphi^\mathbf{J})^{j^*}_{\mathbf{J}(D)}=  \operatorname{sign}(-\langle\kappa,S^{j^*}\rangle \,v_{\mathbf{J}(D)}) = \operatorname{sign}(+2 \cdot -1)  < 0. \\
\operatorname{sign}(\varphi^\mathbf{J})^{j^*}_{\mathbf{J}(E)}=  \operatorname{sign}(-\langle\kappa,S^{j^*}\rangle \,v_{\mathbf{J}(E)}) = \operatorname{sign}(+2 \cdot -1)  < 0.\\
\operatorname{sign}(\varphi^\mathbf{J})^{j^*}_{\mathbf{J}(F)}=  \operatorname{sign}(-\langle\kappa,S^{j^*}\rangle \,v_{\mathbf{J}(F)}) = \operatorname{sign}(+2 \cdot -1)  < 0.\\
\end{cases}
\end{equation}

\subsection{Indeterminate sign Jacobian does not imply indeterminate sign response} \label{indnoind}
 \textcolor{white}{di}

{Definition \ref{detsigndef} introduces the notion of \emph{determinacy} for the response signs. Clearly,}  
the same definition identically applies to the Jacobian determinant  $\operatorname{det}SR$. In the context of sensitivity analysis, a natural question arises for a Jacobian of indeterminate sign, {i.e., when $\operatorname{sign}\operatorname{det}SR$ depends on the values of the derivatives $r_{jm}$}. The question asks whether a Jacobian of indeterminate sign automatically implies that all responses $(\Phi)^{j^*}_{j'}$ are of indeterminate sign. In other words, for
\begin{equation*}
(\Phi)^{j^*}_{j'}=\frac{\sum_{{j^*}\not\in\mathbf{J}\ni{j'} } (\varphi^\mathbf{J})^{j^*}_{j'}}{\operatorname{det}SR},
\end{equation*} 
we ask whether we can exclude that the numerator and denominator switch sign at a same shared root. The answer is negative, cancellations between numerator and denominator may occur, and there may be \emph{determinate sign responses} even in presence of an \emph{indeterminate sign Jacobian}. The following example has been intentionally designed to illustrate such a case. 

\begin{minipage}{.5\textwidth}
\begin{align}
S=
\begin{blockarray}{cccccc}
 & 1 & 2 & 3 & 4& 5\\
\begin{block}{c[ccccc]}
  A & -1 & -1 & -1& 0 & 1 \\
  B & 0 &1 &  1 & -1 & 0\\
  C & 0 &1 & 0 & 1 & -1\\
\end{block}
\end{blockarray}\; , 
\end{align}
\end{minipage}%
\begin{minipage}{.5\textwidth}
\begin{center} 
\includegraphics[scale=0.38]{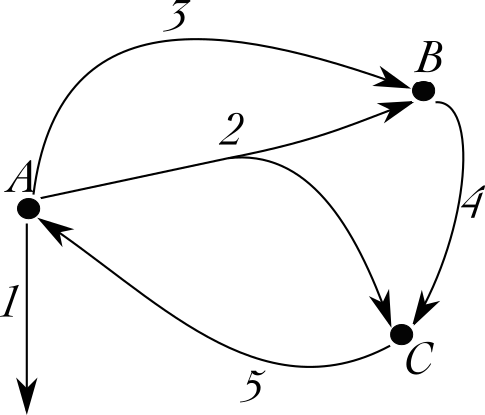}
\end{center}
\end{minipage}

A positive kernel vector of $S$ is $\mathbf{r}=(r,r,r,2r,3r)^T$, $r \in \mathbb{R}_{> 0}$, hence the associated dynamical system admits a positive equilibrium $\bar{x}$ and the network suits our analysis. There are three Child Selections.
\begin{equation}
\begin{cases}
\mathbf{J}_1:= \{\mathbf{J}_1(A)=1; \mathbf{J}_1(B)=4; \mathbf{J}_1(C)=5\}\\
\mathbf{J}_2:= \{\mathbf{J}_2(A)=2; \mathbf{J}_2(B)=4; \mathbf{J}_2(C)=5\}\\
\mathbf{J}_3:= \{\mathbf{J}_3(A)=3; \mathbf{J}_3(B)=4; \mathbf{J}_3(C)=5\}
\end{cases} 
\end{equation}
The sign of the Jacobian determinant is indeterminate. Indeed,
\begin{equation}
\begin{split}
\operatorname{det}SR=&\sum_\mathbf{J} \operatorname{det}S^\mathbf{J} \prod_{m\in \mathbf{M}} r_{\mathbf{J}(m)m}\\
=&\operatorname{det}S^{\mathbf{J}_1}\prod_{m\in \mathbf{M}} r_{\mathbf{J}_1(m)m}+\operatorname{det}S^{\mathbf{J}_2}\prod_{m\in \mathbf{M}} r_{\mathbf{J}_2(m)m}+\operatorname{det}S^{\mathbf{J}_3}\prod_{m\in \mathbf{M}} r_{\mathbf{J}_3(m)m}\\
=&-1\cdot \prod_{m\in \mathbf{M}} r_{\mathbf{J}_1(m)m}+ 1 \cdot \prod_{m\in \mathbf{M}} r_{\mathbf{J}_2(m)m}+0 \cdot \prod_{m\in \mathbf{M}} r_{\mathbf{J}_3(m)m}\\
=& (r_{2A}-r_{1A})r_{4B}r_{5C}.
\end{split}
\end{equation}
At the parameter value $r_{1A}=r_{2A}$, $\operatorname{det}SR$ switches sign. Via Formula \eqref{drjia3}, the flux response $(\Phi)^3_4$ of reaction $4$ to a perturbation of reaction $3$ reads:
\begin{equation}
\begin{split}
(\Phi)^3_4&=\frac{\sum_{4 \in \mathbf{J} \not \ni 3} (\varphi^\mathbf{J})^3_4} {\operatorname{det}SR} =-\frac{(\varphi^{\mathbf{J}_1})^3_4+(\varphi^{\mathbf{J}_2})^3_4}{(r_{2A}-r_{1A})r_{4B}r_{5C}}\\
&=\frac{-\operatorname{det} S^{\mathbf{J}_1 \setminus 4 \cup 3}  \prod_{m\in M} r_{\mathbf{J}_1(m)m} - \operatorname{det} S^{\mathbf{J}_2 \setminus 4 \cup 3}  \prod_{m\in M} r_{\mathbf{J}_2(m)m}}{(r_{2A}-r_{1A})r_{4B}r_{5C}}\\
&=\frac{-1\cdot  \prod_{m\in M} r_{\mathbf{J}_1(m)m} - (-1)  \cdot  \prod_{m\in M} r_{\mathbf{J}_2(m)m}}{(r_{2A}-r_{1A})r_{4B}r_{5C}}\\
&=  \frac{(r_{2A}-r_{1A})r_{4B}r_{5C}}{(r_{2A}-r_{1A})r_{4B}r_{5C}}=1.
\end{split}
\end{equation}
This concludes that $(\Phi)^3_4 \equiv 1$ has determinate sign, with no dependence at all on reaction rates parameters, even if the Jacobian $\operatorname{det}SR$ is of indeterminate sign.\\

\subsection{Subnetwork patterns}\textcolor{white}{di}\\

The results of this paper draw attention to certain subnetworks associated to \emph{Extended Child Selections}  $\mathbf{J} \cup j^*$, for $j^* \not \in \mathbf{J}$, whose associated matrix $S^{\mathbf{J} \cup j^*}$ has a 1-dim kernel. It is of great importance, then, to easily identify such relevant subnetworks.\\ 

Consider firstly Child Selections $\mathbf{J}$ such that
\begin{equation}\label{noz}
\operatorname{det}S^\mathbf{J} \neq 0.
\end{equation}
Condition \eqref{noz} can be characterized on a network level \cite{VGB20}. Then, for $j^*\not \in \mathbf{J}$,
\begin{equation}
\operatorname{dim}\operatorname{ker}S^{\mathbf{J} \cup j^*}=1.
\end{equation}
On the other hand, we lack network characterizations of Child Selections $\mathbf{J}$ such that
\begin{equation}\label{z}
\operatorname{det}S^\mathbf{J} = 0,\quad \quad \quad \text{with} \quad \quad \quad \operatorname{dim}\operatorname{ker}S^{\mathbf{J}}=1,
\end{equation}
which is an obvious necessary condition for
\begin{equation}
\operatorname{dim}\operatorname{ker}S^{\mathbf{J} \cup j^*}=1\quad \quad \quad \text{with $j^*\not \in \mathbf{J}$}.
\end{equation}
One attractive possibility to detect such zero-behaving Child Selections with 1-dim kernels might be to have atomic patterns, in the following sense. Let us assume that we consider only a small, `atomic', part $\mathbf{\Gamma}_{atm}$ of a network $\mathbf{\Gamma}$; for example, a reversible reaction from metabolite $A$ to metabolite $B$. \\

\begin{minipage}{.5\textwidth}
\begin{equation}
S_{atm}=
\begin{blockarray}{ccc}
 & 1 & 2 \\
\begin{block}{c[cc]}
  A & -1 & 1\\
  B & 1 & -1\\
\end{block}
\end{blockarray}\; ,   \;\;
\end{equation}
\end{minipage}%
\begin{minipage}{.5\textwidth}
\begin{center} 
\includegraphics[scale=0.48]{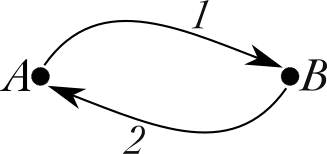}
\end{center}
\end{minipage}

\vspace{0.3cm}

Above, $S_{atm}$ is the stoichiometric matrix of the subnetwork $\mathbf{\Gamma}_{atm}$. Note, in particular, that $\operatorname{dim}(\operatorname{ker} S_{atm})=1$. Let $S$ be the stoichiometric matrix of the whole network $\mathbf{\Gamma}$. Under our standing nondegeneracy assumption $\operatorname{det}SR\neq0$, for the whole network $\mathbf{\Gamma}$, 
\begin{equation}\label{qce}
\begin{split}
\textit{Can we imply the existence of a Child Selection $\mathbf{J}$ for $\mathbf{\Gamma}$},\\ 
\textit{such that} \quad  \mathbf{J}(A)=1,\quad \mathbf{J}(B)=2  \quad \textit{and} \quad \operatorname{dim}(\operatorname{ker} S^\mathbf{J})=1?
\end{split}
\end{equation}

This question possibly has an affirmative answer for most biological networks, so that it may still be a valid strategy in applications. However, on a purely mathematical basis the answer to the question \eqref{qce} is negative. We show this in the following network $\mathbf{\Gamma}$.

\begin{equation} 
\includegraphics[scale=0.38]{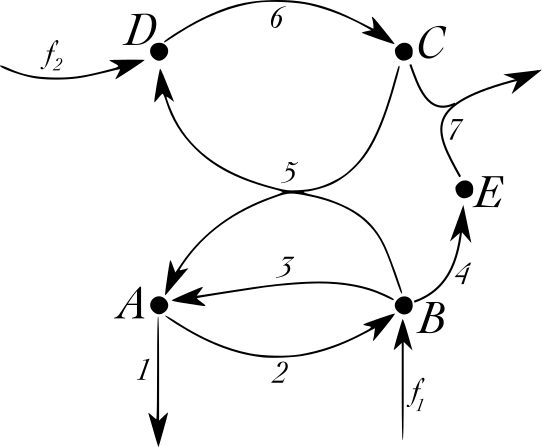}
\end{equation}
\[
S=
\begin{blockarray}{cccccccccc}
 & f_1 &f_2 & 1 & 2 & 3 & 4& 5 & 6 & 7\\
\begin{block}{c[ccccccccc]}
  A & 0 & 0 & -1 & -1 & 1& 0 & 1& 0 & 0\\
  B & 1 & 0 & 0 & 1 &  -1 & -1 & -1 & 0 & 0\\
  C & 0 & 0 & 0 & 0 & 0 & 0 & -1& 1 & -1\\
  D & 0 & 1 & 0 & 0 & 0  & 0 & 1 & -1 & 0\\
  E & 0 & 0 & 0 & 0 & 0 & 1 & 0 & 0 & -1\\
\end{block}
\end{blockarray}\; .
\]
A positive kernel vector of $S$ is $\mathbf{r}=(2r,r,r,r,r,r,r,2r,r)^T$, $r \in \mathbb{R}_{> 0}$, hence the associated dynamical system admits a positive equilibrium $\bar{x}$ and the network suits our analysis. Due to injectivity, there is only one nonzero-behaving Child Selection $\tilde{\mathbf{J}}$, and the nondegeneracy assumption $\operatorname{det}SR \neq 0$ holds. In fact, consider the Child Selection
\begin{equation}
\tilde{\mathbf{J}}:= \{ \tilde{\mathbf{J}}(A)=1; \tilde{\mathbf{J}}(B)=4; \tilde{\mathbf{J}}(C)=5; \tilde{\mathbf{J}}(D)=6; \tilde{\mathbf{J}}(E)=7\},
\end{equation}
with associated nonsingular stoichiometric matrix
\begin{align}
S^{\tilde{\mathbf{J}}}=
\begin{blockarray}{cccccc}
 & 1 & 4 & 5 & 6 & 7\\
\begin{block}{c[ccccc]}
  A & -1 & 0& 1& 0 & 0\\
  B & 0 &  -1 & -1 & 0 & 0\\
  C & 0 & 0 & -1& 1 & -1\\
  D & 0 & 0  & 1 & -1 & 0\\
  E & 0 & 1 & 0 & 0 & -1\\
\end{block}
\end{blockarray}\;,\quad
\operatorname{det}S^{\tilde{\mathbf{J}}}=1,
\end{align}
hence
$$\operatorname{det}SR=\operatorname{det}S^{\tilde{\mathbf{J}}}\prod_{m\in \mathbf{M}} r_{\tilde{\mathbf{J}}(m)m}=\prod_{m\in \mathbf{M}} r_{\tilde{\mathbf{J}}(m)m} \neq 0.$$ 
Reactions $2$ and $3$ and their input metabolites $A$ and $B$ constitute a degenerate subnetwork $\mathbf{\Gamma}_{atm} \subset \mathbf{\Gamma}$ whose stoichiometric matrix $S_{atm}$ reads:
\begin{align}
S_{atm}=
\begin{blockarray}{ccc}
 & 2 & 3 \\
\begin{block}{c[cc]}
  A & -1 & 1\\
  B & 1 &  -1\\
\end{block}
\end{blockarray}\;.
\end{align}
We show that it is not possible to extend the Child Selection $\{ {\mathbf{J}}_{atm}(A)=2,\; {\mathbf{J}}_{atm}(B)=3 \}$ on $\mathbf{\Gamma}_{atm}$ to a Child Selection ${\mathbf{J}_{23}}$ on $\mathbf{\Gamma}$ such that $\operatorname{dim}(\operatorname{ker}S^{{\mathbf{J}_{23}}})=1$. {Indeed, there is only one possible Child Selection $\mathbf{J}_{23}$ of $\Gamma$ with $2, 3 \in \mathbf{J}_{23}$, that is:
\begin{equation}
{\mathbf{J}_{23}}:= \{{\mathbf{J}_{23}}(A)=2; {\mathbf{J}_{23}}(B)=3; {\mathbf{J}_{23}}(C)=5; {\mathbf{J}_{23}}(D)=6; {\mathbf{J}_{23}}(E)=7\},
\end{equation}
with associated stoichiometric matrix
\begin{align}
S^{{\mathbf{J}_{23}}}=
\begin{blockarray}{cccccc}
 & 2 & 3 & 5 & 6 & 7\\
\begin{block}{c[ccccc]}
  A & -1 & 1& 1& 0 & 0\\
  B & 1 &  -1 & -1 & 0 & 0\\
  C & 0 & 0 & -1& 1 & -1\\
  D & 0 & 0  & 1 & -1 & 0\\
  E & 0 & 0 & 0 & 0 & -1\\
\end{block}
\end{blockarray}\;,
\end{align}
which possesses a 2-dimensional kernel, $\operatorname{ker}(S^{{\mathbf{J}_{23}}})=\operatorname{span}\langle v^1,v^2\rangle$ where $v^1=(1,1,0,0,0)^T$ and $v^2=(1,0,1,1,0)^T$.}

\section{Metabolite perturbation}\label{metpertsec}

{
It may be of interest for the reader a brief discussion on \emph{metabolite perturbation}, rather than \emph{reaction perturbation}. This case is extendedly treated in a dedicated paper \cite{V21}, where more details can be found. This section is intended only as a brief overview on the topic, with the aim of keeping this paper as self-contained as possible.\\
Within our setting, a natural possibility to include metabolite perturbations of equilibria is the following perturbed equation (cf: \eqref{perteq}):
\begin{equation}\label{metpert}
0=S\mathbf{r}(\bar{x})+\varepsilon e_{m^*}.
\end{equation}
Equation \eqref{metpert} perturbs the metabolite $m^*$ by adding a constant inflow to it. Such perturbation has been considered for example by the ecology community \cite{Yo88,NA92}, with the name of \emph{`press experiments'}. Under the standing assumption $\operatorname{det}SR \neq 0$ and along the lines of Section \ref{sensitivity}, we may investigate the response $\delta x^{m^*}_{m'}$ of the concentration of the metabolite $m'$ and the response $\Phi^{m^*}_{j'}$ of the flux of the reaction $j'$. We obtain the following relations:
\begin{equation}\label{metresponses}
\delta x^{m^*}_{m'}= -((SR)^{-1}e_{m^*})_{m'} \quad \quad \text{and} \quad \quad \Phi^{m^*}_{j'}=-[R(SR)^{-1} e_{m^*}]_{j'}.
\end{equation}
Let $p'$ indicate $m'$ or $j'$, indistinctly. A nonzero response of $p'$ is called \emph{nonzero influence} of $m^*$ on $p'$ and indicated by
$$m^* \quad \rightsquigarrow \quad p' .$$
Note that the above equation can be interpreted as a reaction perturbation of an inflow reaction to $m^*$. This is confirmed by the parallelism between responses \eqref{fluxmetresp}, \eqref{fluxfluxresp}, and \eqref{metresponses}. Thus, mathematically, this type of metabolite perturbation is a subcase of the more general reaction perturbation. However, it is worth mentioning two counterintuitive features of this case. Firstly, the response of $m^*$ may be zero upon a perturbation of $m^*$ itself. That is, adding an $\varepsilon$-inflow to $m^*$, the new equilibrium may have the same concentration of metabolite $m^*$, thus self-influence does not always happen:
$$m^* \quad \not\rightsquigarrow \quad m^*.$$
Secondly, transitivity of influence does not hold for this case of influence:
$$m_1  \rightsquigarrow  m_2   \rightsquigarrow  m_3 \quad\quad \not\Rightarrow \quad\quad m_1   \rightsquigarrow  m_3.$$
The transitivity failure is not due to a discrepancy between reaction perturbation and metabolite perturbation, of course: the failure is only due to the different algebraic structure of the metabolite response $\delta x_m$ compared to the flux response $\Phi_j$.\\
Explicit examples of both features can be found in \cite{V21} and we omit them here.}

\section{Discussion} \label{discussion}

We have addressed the sign of {flux sensitivities} to perturbations of the network components and identified certain kernel vectors of the stoichiometric matrix, which encode the signs. These kernel vectors are constructed by considering a Child Selection $\mathbf{J}$ and a reaction $j^*$ such that $j^* \not \in \mathbf{J}$. Each Extended Child Selection (ECS) $\mathbf{J} \cup j^*$ identifies a $M \times (M+1)$ minor $S^{\mathbf{J}\cup j^*}$ of the stoichiometric matrix. When such minor has a 1-dim kernel, its kernel is spanned by a single vector $v \in \mathbb{R}^{M+1}$,
$$\operatorname{ker}S^{\mathbf{J} \cup j^*}=\operatorname{span}\langle v \rangle.$$
These kernels $v$, which we have named ECS kernel vectors, are the precise network structures encoding the signs of the responses, in the sense explained in Section \ref{SRMR}. Note that, for any matrix with 1-dim kernel, the support, i.e. nonzero entries, of any nontrivial kernel vector is uniquely defined. This implies, trivially, that the support of such kernel vectors does not properly contain the support of any other kernel vector. Such property defines what in literature has been named \emph{elementary} kernel vectors, mathematically studied by Rockafellar \cite{Rock69}. Note also that to each ECS vector $v$ we can naturally associate a unique kernel vector $v^S \in \mathbb{R}^N$ of the full stoichiometric matrix $S$ by considering
\begin{equation}
v^S_j=
\begin{cases}
v_j\quad \;\;\; \text{if}\quad j\in \mathbf{J}\cup j^*,\\
0\quad \quad \text{otherwise}.
\end{cases}
\end{equation}
In this sense, ECS vectors are in particular special \emph{elementary kernel vectors} of the stoichiometric matrix $S$. Interestingly, the importance of elementary kernel vectors of the stoichiometric matrix in metabolic networks has already been noted. Important concepts as \emph{elementary flux modes} \cite{SchHil94} and \emph{elementary flux vectors} \cite{Kl17} have arisen in a different metabolic context from the one treated in the present paper, with connections yet to be investigated in detail.\\
In this paper we have not addressed computational issues related to computing such ECS kernel vectors. {The number of Child Selections $n_\mathbf{J}$ can be estimated:
\begin{equation}
n_\mathbf{J} \; \le \; \prod_{m} n_{J(m)},
\end{equation}
where $n_{J(m)}$ is the number of reactions $j$ of which $m$ is an input metabolite. The inequality is due to the injectivity assumption in Definition \ref{CSDEF}, and the equality is thus obtained only if no two metabolites $m_1$ and $m_2$ are input of a same reaction $j$; that is, each reaction in the network possesses only one metabolite input, e.g. in monomolecular networks. In particular, $n_\mathbf{J}$ grows exponentially with the number of the metabolites $m$. However,} an algorithm considering each Child Selection $\mathbf{J}$ and each reaction $j^* \not \in \mathbf{J}$ may not be the most efficient. In fact, as we found in Example \ref{Signtrans}, \eqref{smk}, to {two} different Child Selections $\mathbf{J}_1, \mathbf{J}_2$, and a reaction $j^* \not \in \mathbf{J}_1, \mathbf{J}_2$, may correspond the \emph{same} single ECS kernel vector. In other words, there may be much fewer ECS kernel vectors than Extended Child Selections. An efficient computing algorithm would be of great help in making the results of the present paper more operative for daily analysis on real biological networks.\\

Example \ref{Signtrans} presented a simple counterexample to sign-transitivity of influence. With such result, the question of transitivity of influence in metabolic networks is answered. We summarize it here, for sake of clarity. Firstly, we recall a positive result of \cite{BF18} on the topic:
\begin{thm}[Brehm-Fiedler] \label{BFtrans}
Let $p_1$ and $p_2$ be elements in a metabolic network, either metabolites or reactions. Let $j'$ be any reaction and $m'$ one of its input metabolites.
\begin{enumerate}
\item If $p_1 \rightsquigarrow m'$ and $j'\rightsquigarrow p_2$,  \quad \, then $p_1\rightsquigarrow p_2$. 
\item If $p_1\rightsquigarrow j'$ and $j' \rightsquigarrow p_2$, \quad \; then $p_1 \rightsquigarrow p_2$.
\end{enumerate}
\end{thm}
In \cite{V21}, it is proven that the result \ref{BFtrans} does \emph{not} extend to the metabolite case. That is, 
$$ p_1 \rightsquigarrow m' \text{ and } m'\rightsquigarrow p_2,  \quad \not\Rightarrow \quad p_1\rightsquigarrow p_2,$$
and the present paper shows that no sign-transitivity result hold. That is,
$$p_1 \mathrel{\mathop{\rightsquigarrow}^{\mathrm{+}}} p_2 \mathrel{\mathop{\rightsquigarrow}^{\mathrm{+}}} p_3 \not\Rightarrow  p_1 \mathrel{\mathop{\rightsquigarrow}^{\mathrm{+}}} p_3,$$
or any other combination of sign. In conclusion, Theorem \ref{BFtrans} covers all transitivity properties and no more general result holds.\\

{We started this paper with three questions: we discuss here answers. Theorem \ref{signres1} indicates which reaction $j^*$ should be perturbed to achieve an influence on $j'$: at least one of the ECS kernel vectors $v$ associated to $j^*$ must have a nonzero $j'$-th entry, $v_j' \neq 0$. For the control of the response sign, we must distinguish determinate or indeterminate sign. If \emph{determinate}, the sign of a response is robust: it is independent from the equilibrium value $\bar{x}$ and from any chosen kinetics. In particular, the sign is the same for all choices of reaction rates. The only naive way to control a nonzero sign is changing the perturbation itself. For instance, Section \ref{twins} describes the lucky case in which any Child Selection maps a metabolite $m^*$ either to $j^*$ or $j^*_s$, only: the flux-responses to a perturbation of $j^*$ have always opposite sign to the responses to a perturbation of $j^*_s$. Hence, for a \emph{positive} influence on the flux of $j'$, we may just choose between perturbing $j^*$ or $j^*_s$. Alternatively, we may consider a sign switch of the perturbation. Indeed, consider a \emph{negative} perturbation of $\tilde{\varepsilon}=-\varepsilon \le 0$: by linearity, the responses to a $\tilde{\varepsilon}$-perturbation have opposite sign of the responses to a $\varepsilon$-perturbation.\\ 
More interestingly, for a fixed perturbation, if a response $(\Phi)^{j^*}_{j'}$ is of \emph{indeterminate} sign, different values of the reaction rates may produce different signs of $(\Phi)^{j^*}_{j'}$. In this sense, the control of the response sign may be possible via a careful choice of the reaction rates, alone. However, this does strongly depend on the class of nonlinearities $\mathbf{r}$, the \emph{kinetics}.} Metabolic networks are usually endowed with \emph{enzymatic kinetics} as - for example - Michaelis-Menten
\begin{equation}\label{michment}
r_j(x)=a_j\prod_{m\in\mathbf{M}} \Bigg( \frac{x_m}{(1+b_j^m x_m)}\Bigg)^{s^j_m},
\end{equation}
where $s^j_m$ is the stoichiometric coefficient of metabolite $m$ in the reaction $j$, and $a_j$ and $b_j^m$ are positive parameters. The reason for this choice is that the reactions appearing in the network only describe relations between metabolites. However, also other chemicals may be involved, in particular enzymes. The presence of enzymes is taken in account not by the stoichiometry of the network but indeed by choosing enzymatic kinetics rather than an elementary kinetics, such as mass action
\begin{equation}
r_j(x)=c_j \prod_{m\in\mathbf{M}} x_m^{s^j_m},
\end{equation}
where again $s^j_m$ is the stoichiometric coefficient of metabolite $m$ in the reaction $j$, and $c_j$ is \emph{one single} positive parameter. A mathematical advantage of Michaelis-Menten over mass action is the greater richness of parameters, as \emph{$\#$inputs}+1 parameters appear in the rate of any reaction $j$, where \emph{$\#$inputs} indicates the number of input metabolites to reaction $j$. This {parametric richness has important consequences for the control of the sign of the responses. In fact, in Michaelis-Menten \eqref{michment}, a careful choice of the positive parameters $a_j$ and $b_j^m$ enables us to} consider, at any fixed equilibrium $\bar{x}$, the derivatives $r_{jm}(\bar{x})$ as free positive parameters, independent from each other and from the equilibrium $\bar{x}$ itself. We refer again to \cite{VGB20} for an explicit computation of this mathematical fact. For what concerns the present paper, in the case of a kinetics as parametrically rich as Michaelis-Menten, there always exist choices of reaction rate parameters so that a response of indeterminate sign {can be controlled} to be positive, negative, or zero. For parametrically poorer kinetics, as mass action, this freedom is missing: only one single parameter $c_j$ appears in the rate of each reaction $j$. Thus, for mass action, further analysis must be performed to fully understand and possibly control the actual sign of a response of indeterminate sign. {A viable and valid strategy may be carefully choosing also the equilibrium value $\bar{x}$, oppositely to the approach presented in this paper, where we have considered $\bar{x}$ fixed. However, consider a metabolite $m$ that is input to two reactions $j_1$ and $j_2$. Clearly, an `equilibrium parameter' $\bar{x}_m$ may appear in both the mathematical expressions of the derivatives $r_{j_1m}$ and $r_{j_2m}$. Contrarily to our approach then, if the equilibrium $\bar{x}$ itself is treated as a parameter, we may not consider $r_{j_1m}$ and $r_{j_2m}$ as parametrically independent, even in the Michaelis-Menten case. This indeed requires further analysis, untouched by the present work.}\\

Throughout the paper, we have assumed the nondegeneracy assumption \eqref{nondeg} $$\operatorname{det}SR\neq 0,$$
where $SR$ is the Jacobian matrix of the system, which excludes left kernels (conserved quantities) of the stoichiometric matrix $S$. {We do not exclude that it may be mathematically possible to relax \eqref{nondeg} to include such case. We have not pursued this in the present paper both to a greater mathematical clarity of the content and because the assumption \eqref{nondeg} already allows important examples in a metabolic context, due to the omnipresence of outflows.}\\
Based on the implicit function theorem, the present theory appears firstly as a local theory, valid only for small perturbations. However, an interpolation argument proposed in \cite{BF18} lifts formula \eqref{drjia3}, and consequently these sensitivity results, to account also for large perturbations, in the case of zero-vs-nonzero response analysis. {Of course, a further assumption on the existence of an equilibrium of the largely perturbed system must be added. Then the flux responses to a large perturbation follow the same algebraic description as in the local case. In particular,} the interpolation argument works identically also for the sign analysis presented in the present paper. However, carefulness is strongly required here, {as we can never choose the parameters independently from the equilibrium, as it is possible locally for parametrically rich kinetics. This implies that,} if the sign of a response is determinate, i.e., not depending on parameters, then the sign of the response does {not} depend on the amplitude of the perturbation. On the contrary, in the case of a response of indeterminate sign, i.e., depending on parameters, a positive response for a small perturbation may become a negative response for a large perturbation, even for parametrically rich kinetics as Michaelis-Menten.\\

In applications, the system is often considered to have a unique stable equilibrium, for any choice of reaction rates. {Under the nondegeneracy assumption \eqref{nondeg}}, this requires that the Jacobian determinant $\operatorname{det}SR$ is of the \emph{determinate} sign
\begin{equation}\label{detsign}
\operatorname{sign}\operatorname{det}SR=(-1)^M,
\end{equation}
since all $M$ eigenvalues {have negative real part, or are purely imaginary complex conjugated pairs}. Via expansion \eqref{detexp}, i.e.,
$$\operatorname{det}SR= \sum_{\mathbf{J}} \operatorname{det} S^\mathbf{J} \cdot \prod_{m \in \mathbf{M}} r_{\mathbf{J}(m)m},$$
it is easy to see that an obvious sufficient condition for \eqref{detsign} is that all Child Selections are \emph{good}, that is
$$\beta(\mathbf{J})=\operatorname{sign}\operatorname{det}S^\mathbf{J}\equiv(-1)^M, \quad \quad \quad \text{for any } \mathbf{J}.$$
In particular, trivially, $\operatorname{sign}\operatorname{det}SR=\beta(\mathbf{J})$, for any $\mathbf{J}$. For such special but relevant case, we can express better the sign of the response $(\Phi)^{j^*}_{j'}$. Indeed, via \eqref{drjia3} we have 
\begin{equation}
\operatorname{sign}(\Phi)^{j^*}_{j'}= \frac{\operatorname{sign}(\sum_{{j^*}\not\in\mathbf{J}\ni{j'}} (\varphi^\mathbf{J})^{j^*}_{j'})}{\operatorname{sign}\operatorname{det}SR}=\beta(\mathbf{J})\operatorname{sign}\bigg(\sum_{{j^*}\not\in\mathbf{J}\ni{j'}}(\varphi^\mathbf{J})^{j^*}_{j'}\bigg),
\end{equation}
but, via \eqref{drjia}, for each $\mathbf{J}$, such that ${j^*}\not\in\mathbf{J}\ni{j'}$,
\begin{equation}
\beta(\mathbf{J}) \operatorname{sign} (\varphi^\mathbf{J})^{j^*}_{j'} = \beta(\mathbf{J}) \beta(\mathbf{J})\operatorname{sign}(v^\mathbf{J^*}_{j^*}v^\mathbf{J^*}_{j'})=\operatorname{sign}(v^\mathbf{J^*}_{j^*}v^\mathbf{J^*}_{j'}),
\end{equation}
where the notation $v^\mathbf{J^*}$ indicates here the usual ECS kernel vector such that
$$S^{\mathbf{J}\cup j^*}v^{\mathbf{J}^*} = 0.$$
Clearly, then, the response $(\Phi)^{j^*}_{j'}$ is of indeterminate sign \emph{if and only if} there are two Child Selections $\mathbf{J}_1$ and $\mathbf{J}_2$, such that $j^* \not \in \mathbf{J}_1, \mathbf{J}_2$, $j' \in \mathbf{J}_1, \mathbf{J}_2$, and the associated ECS kernel vectors $v^\mathbf{J^*_1}$ and $v^\mathbf{J^*_1}$ are such that
\begin{equation}
v^{\mathbf{J}_1^*}_{j^*}v^{\mathbf{J}_2^*}_{j'} \;<\; 0\; <\;  v^{\mathbf{J}_2^*}_{j^*}v^{\mathbf{J}_2^*}_{j'}.
\end{equation}\\

Interesting questions still arise in the much more challenging case of an indeterminate sign Jacobian. On the one hand, we have showed in Example \ref{indnoind} that cancellations may occur between the numerator and the denominator of 
\begin{equation*}
(\Phi)^{j^*}_{j'}=\frac{\sum_{{j^*}\not\in\mathbf{J}\ni{j'} } (\varphi^\mathbf{J})^{j^*}_{j'}}{\operatorname{det}SR},
\end{equation*} 
so that an indeterminate sign of the Jacobian determinant does not a priori imply the indeterminate sign of the responses. On the other hand, the example is artificially constructed and this quite surprising feature may not often happen in real biological networks. Addressing and characterizing in more detail network conditions leading to such cancellations is of great interest for future work: if cancellations are excluded, the sign of all responses undergoes a \emph{simultaneous switch} at the zero of the Jacobian, at the same bifurcation point of a possible saddle-node bifurcation, connecting the control of the sign of the sensitivity responses to stability properties of the equilibrium.\\

\section{Proofs} \label{Signproof}

We start this section with the proof of Proposition \ref{basic}.

\proof[Proof of Proposition \ref{basic}]

Preliminarily, note that $\operatorname{ker}(S^{\mathbf{J}\cup j^*}) \neq \emptyset $, since $S^{\mathbf{J}\cup j^*}$ is a $M \times (M+1)$ matrix. Hence, the dimension of the kernel is either $1$ or greater than $1$. Moreover, by Formula \eqref{drjia2}, 
\begin{equation}
(\varphi^\mathbf{J})^{j^*}_{j'} \neq 0 \quad \Leftrightarrow \quad \operatorname{det}(S^{\mathbf{J}\setminus j' \cup j^*})\neq 0.
\end{equation}

Firstly, assume that $\operatorname{dim}(\operatorname{ker}(S^{\mathbf{J}\cup j^*}))>1$. 
\begin{equation}
\begin{split}
\operatorname{dim}(\operatorname{ker}(S^{\mathbf{J}\cup j^*}))>1 \quad &\Rightarrow \quad \operatorname{ker}(S^{\mathbf{J}\setminus j' \cup j^*}) \neq \emptyset,\text{ for all $j' {\in\mathbf{J}}$}  \\ 
&\Rightarrow \quad (\varphi^\mathbf{J})^{j^*}_{j'} = 0,\text{ for all $j' {\in \mathbf{J}}$}.
\end{split}
\end{equation}

Conversely, assume that $\operatorname{dim}(\operatorname{ker}(S^{\mathbf{J}\cup j^*}))=1$. We have
\begin{equation}
\begin{split}
\operatorname{dim}(\operatorname{ker}(S^{\mathbf{J}\cup j^*}))=1 \quad &\Rightarrow \quad \operatorname{rank} S^{\mathbf{J}\cup j^*}= M\\ & \Rightarrow \quad \exists \; \operatorname{det}(S^{\mathbf{J}\setminus j' \cup j^*})\neq 0 \quad  \Rightarrow \quad { \exists \; j' \text{ such that}} \;(\varphi^\mathbf{J})^{j^*}_{j'} \neq 0.
\end{split}
\end{equation}

\endproof

\proof[Proof of Theorem \ref{signres1}] The proof is based on a careful use of Cramer's rule.\\

1) We prove that
\begin{equation}
(\varphi^\mathbf{J})^{j^*}_{j'}\neq0 \quad \Leftrightarrow \quad v_{j'}\neq 0.
\end{equation}
The first step is to make the matrix $S^{\mathbf{J}\cup j^*}$ an invertible $(M+1)\times(M+1)$ matrix $N_b$ by adding in the $(M+1)$-th row a proper row vector $b^T$, that is
\begin{align}
N_b:=
\begin{bmatrix}
S^{\mathbf{J}\cup j^*}\\
b^T\\
\end{bmatrix}\; .   
\end{align}
Secondly, we compute:
\begin{align}
\begin{bmatrix}
S^{\mathbf{J}\cup j^*}\\
b^T\\
\end{bmatrix} \cdot {v} =
\begin{bmatrix}
\underline{0}\\
\langle b,v \rangle\\
\end{bmatrix} .
\end{align}
Above, $\underline{0}$ refers to the $M$-dimensional zero vector. Note that $\langle b,v \rangle \neq 0$, since $N_b$ is invertible. We now apply Cramer's rule to the $j'$-th entry of $v$ and find that
\begin{align} \label{kern}
\begin{split}
\operatorname{det}(N_b) \; v_{j'}&=
\operatorname{det}
\begin{blockarray}{ccccc}
1 & ... & j' & ... & M+1\\
\begin{block}{[ccccc]}
S^{\mathbf{J}(m_1)} & ... & \underline{0} & ... & S^{j^*}\\
b^T_1 & ... & \langle b,v \rangle & ... & b^T_{M+1}\\
\end{block}
\end{blockarray}\\
&=
-\operatorname{det}
\begin{blockarray}{ccccc}
1 & ... & j' & ... & M+1\\
\begin{block}{[ccccc]}
S^{\mathbf{J}(m_1)} & ... & S^{j^*} & ... & \underline{0} \\
b^T_1 & ... & b^T_{M+1} & ... & \langle b,v\rangle\\
\end{block}
\end{blockarray}\\
&= - \langle b,v\rangle \, \operatorname{det} S^{\mathbf{J}\setminus j' \cup j^*}.
\end{split}
\end{align}
The conclusion follows by noting that 
\begin{equation}
v_{j'} \neq 0 \Leftrightarrow \operatorname{det} S^{\mathbf{J}\setminus j' \cup j^*}\neq 0 \Leftrightarrow (\varphi^\mathbf{J})^{j^*}_{j'}\neq 0.
\end{equation}

2) {Equality \eqref{kern}, in particular, holds for any two $v_{j_1'}, v_{j_2'}\neq 0$, with $j_1',j_2' \in \mathbf{J}$}. We can divide one equality by the other obtaining
\begin{equation}
\frac{v_{j_1'}}{v_{j_2'}}=\frac{\operatorname{det} S^{\mathbf{J}\setminus j_1' \cup j^*}}{\operatorname{det} S^{\mathbf{J}\setminus j_2' \cup j^*}}=\frac{(\varphi^\mathbf{J})^{j^*}_{j'_1}}{(\varphi^\mathbf{J})^{j^*}_{j'_2}}.
\end{equation}

Passing to the sign operator gives the desired equality.
\endproof

\proof[Proof of Theorem \ref{signres2}]

Firstly, let us observe that, under the one-dimensional condition $\operatorname{ker} S^{\mathbf{J} \cup j^*}= \operatorname{span} \langle v \rangle$, we have
\begin{equation}
\operatorname{det} S^\mathbf{J} = 0 \quad \Leftrightarrow \quad v_{j^*}=0.
\end{equation}

1) Now, let us assume $\operatorname{det}  S^\mathbf{J} \neq 0$, i.e. $v_{j^*}\neq 0$. By Cramer's rule,
\begin{align} \label{kernj*}
\begin{split}
\operatorname{det}(N_b) \; v_{j*}&=
\operatorname{det}
\begin{blockarray}{ccc}
1 & ... &  M+1\\
\begin{block}{[ccc]}
S^{\mathbf{J}(m_1)} & ... & \underline{0}\\
b^T_1 & ...  & \langle b,v\rangle \\
\end{block}
\end{blockarray}\\
&= \langle b,v \rangle \operatorname{det} S^{\mathbf{J}}.
\end{split}
\end{align}
Comparison of the equalities between \eqref{kern} regarding ${v_{j'}}$ and \eqref{kernj*} regarding ${v_{j^*}}$ implies:
\begin{equation}
\frac{v_{j'}}{v_{j^*}}=\frac{-\operatorname{det} S^{\mathbf{J}\setminus j' \cup j^*}}{\operatorname{det} S^{\mathbf{J}}}.
\end{equation}
Passing to the sign operator yields
\begin{equation}
\operatorname{sign}(v_{j'} v_{j^*})=\beta(\mathbf{J}) \operatorname{sign} (\varphi^\mathbf{J})^{j^*}_{j'}.
\end{equation}

2)  {The case of $(\varphi^\mathbf{J})^{j^*}_{j'}=0$ is trivially proven. Indeed, by Theorem \ref{signres1} case (1),$$(\varphi^\mathbf{J})^{j^*}_{j'}=0 \Leftrightarrow v_{j'}=0,$$ and thus \eqref{rev2} holds. Assume then $(\varphi^\mathbf{J})^{j^*}_{j'}\neq 0$, and consider the $(M+1)\times(M+1)$ matrix
\begin{align}
N_{j'}:=
\begin{bmatrix}
S^{\mathbf{J}\cup j^*}\\
e_{j'}^T\\
\end{bmatrix}\;,
\end{align}
where $e_{j'}$ indicates the $j'$-th unit vector in $\mathbb{R}^{M+1}$. Note that
\begin{align}
\operatorname{det}(N_{j'}) = (-1)^{j'+M+1}(-1)^{M-j'} \operatorname{det} (S^{\mathbf{J} \setminus j' \cup j^*})=-\operatorname{det} (S^{\mathbf{J} \setminus j' \cup j^*}).
\end{align} }
Hence,
\begin{equation}
0\neq \operatorname{sign}(\varphi^\mathbf{J})^{j^*}_{j'}=-\operatorname{sign}\operatorname{det} (S^{\mathbf{J} \setminus j' \cup j^*})=\operatorname{sign}\operatorname{det}(N_{j'}).
\end{equation}
To compute $\operatorname{det}(N_{j'})$, we consider
\begin{align}
\operatorname{det}(N_{j'})=\operatorname{det}(N_{j'}^T)=\operatorname{det}
\begin{bmatrix}
(S^{\mathbf{J}})^T & e_{j'}\\
 (S^{j^*})^T & 0    \\
\end{bmatrix}.
\end{align}
Let us consider $\tilde{v} \in \mathbb{R}^M$ such that $\operatorname{ker}S^\mathbf{J}=\operatorname{span}\langle \tilde{v}\rangle$ and $\tilde{v}_j = v_j$, for any $j=1,...,M$.\\ 
Now, for square matrices, $\operatorname{dim}\operatorname{coker}(S^{\mathbf{J}})=\operatorname{dim}\operatorname{ker}(S^{\mathbf{J}})$. Let us choose the vector $\kappa \in \mathbb{R}^M$ such that,
\begin{equation}
{\operatorname{ker}(S^{\mathbf{J}})^T}=\operatorname{span}\langle\kappa\rangle,
\end{equation}
and
\begin{equation}
\operatorname{Ad}(S^\mathbf{J}) = \; \tilde{v} \cdot \kappa^T
\end{equation}
Let us set $\tilde{\kappa}=(\kappa, 0)^T$. Again:
\begin{align}
N_{j'}^T \cdot \tilde{\kappa} =
\begin{bmatrix}
\underline{0}\\
\langle S^{j^*},\kappa \rangle\\
\end{bmatrix}.
\end{align}
Let us pick an entry $\kappa_i \neq 0$ and, one more time by Cramer's, we obtain:
\begin{align} \label{coker}
\begin{split}
\operatorname{det}({N_{j'}^T}) \; \kappa_i&=
\operatorname{det}
\begin{blockarray}{ccccc}
  & i & &  M+1\\
\begin{block}{[cccc]c}
 ... & 0 & ...& 0 & 1\\
 ...& ...  & ...& ... & \\
 ... & 0 & ... & 1_{j'} & j'\\
 ...& ...  & ...& ... & \\
 ... & \langle S^{j^*},\kappa\rangle & ... & 0 &  M+1 \\
\end{block}
\end{blockarray}\\
&=
(-1)^{i+j'+1} \langle S^{j^*},\kappa\rangle \; \operatorname{det}(S^\mathbf{J})^{\vee{j'}}_{\vee{i}}.
\end{split}
\end{align}
Above, again, $(S^\mathbf{J})^{\vee{j'}}_{\vee{i}}$ indicates the matrix with removed column $j'$ and row $i$.\\
Now, noting that
\begin{equation} \label{coker2}
(-1)^{i+j'} \operatorname{det}(S^\mathbf{J})^{\vee{j'}}_{\vee{i}} = (\operatorname{Ad}S^\mathbf{J})^{j'}_{i}= v_{j'} \kappa_i 
\end{equation}
leads to the complete chain of equalities:
\begin{equation}
\begin{split}
\operatorname{sign}(\varphi^\mathbf{J})^{j^*}_{j'}&=-\operatorname{sign}\operatorname{det} (S^{\mathbf{J} \setminus j' \cup j^*})=\operatorname{sign}\operatorname{det}(N_{j'})=\operatorname{sign}\operatorname{det}(N_{j'}{^T})\\
&= -\operatorname{sign} ( v_j \langle S^{j^*},\kappa\rangle),
\end{split}
\end{equation}
which concludes our proof.
\endproof

\proof[Proof of Proposition \ref{TwinSisters}]

We have assumed that any Child Selection $\mathbf{J}$ contains either $j^*$ or $j^*_s$, as a child reaction of $m^*$.\\

Let us pick the influence of $j^*$ on any $j'\neq j^*, j^*_s$. Via Formula \eqref{drjia3}:
\begin{equation}
\begin{split}
\operatorname{det}SR \cdot r_{j^*m^*} \cdot (\Phi)^{j^*}_{j'}=& - \sum_{{j^*}\not\in \mathbf{J} \ni {j'}}\operatorname{det}S^{\mathbf{J} \smallsetminus {j'} \cup {j^*}}\cdot r_{j^*m^*} \cdot \prod_{m\in M}r_{\mathbf{J}(m)m} \\
=& + \sum_{{j_s^*}\not\in \tilde{\mathbf{J}} \ni {j'}}\operatorname{det}S^{\tilde{\mathbf{J}} \smallsetminus {j'} \cup {j_s^*}} \cdot r_{j^*_sm^*} \prod_{m\in M}r_{\tilde{\mathbf{J}}(m)m}\\
=&-\operatorname{det}SR \cdot r_{j^*_sm^*} \cdot (\Phi)^{j_s^*}_{j'}.
\end{split}
\end{equation}
To check the central step above, note that any Child Selection, which does not contain $j^*$, must contain $j^*_s$, {the `twin sister' of $j^*$}. Hence, with only one column swap $j^* \leftrightarrow j^*_s$, the matrix $S^{\mathbf{J} \smallsetminus {j'} \cup {j^*}}$, for a Child Selection $\mathbf{J}\not \ni j^*$ becomes the matrix $S^{\tilde{\mathbf{J}} \smallsetminus {j'} \cup {j_s^*}}$ for a Child Selection $\tilde{\mathbf{J}}\not \ni j^*_s$. The step follows since the determinant is an alternating form.\\
Cases  $j^*=j'$ and $j^*_s=j'$ follow analogously by considering Formula \eqref{Bsfluxfluxself} instead. We omit the computation here.
\endproof

\textbf{Acknowledgments:} \textit{This work has been supported by the Collaborative Research Center (SFB) 910 and the Berlin Mathematical School.}\\

\textbf{Data Availability Statement:} \textit{Data sharing not applicable to this article as no datasets were generated or analyzed during the current study.}

 \bibliographystyle{ieeetr}

\begin{thebibliography}{10}

\bibitem{Lopez2008}
L.~L{\'o}pez-Maury, S.~Marguerat, and J.~B{\"a}hler, ``Tuning gene expression
  to changing environments: from rapid responses to evolutionary adaptation,''
  {\em Nature Reviews Genetics}, vol.~9, no.~8, pp.~583--593, 2008.

\bibitem{Zapetal21}
R.~Zappasodi, I.~Serganova, I.~J. Cohen, M.~Maeda, M.~Shindo, Y.~Senbabaoglu,
  M.~J. Watson, A.~Leftin, R.~Maniyar, S.~Verma, {\em et~al.}, ``C{T}{L}{A}-4
  blockade drives loss of $\text{{T}}_{\text{reg}}$ stability in glycolysis-low
  tumours,'' {\em Nature}, vol.~591, no.~7851, pp.~652--658, 2021.

\bibitem{Ish07}
N.~Ishii, K.~Nakahigashi, T.~Baba, M.~Robert, T.~Soga, A.~Kanai, T.~Hirasawa,
  M.~Naba, K.~Hirai, A.~Hoque, {\em et~al.}, ``Multiple high-throughput
  analyses monitor the response of e. coli to perturbations,'' {\em Science},
  vol.~316, no.~5824, pp.~593--597, 2007.

\bibitem{Yo88}
P.~Yodzis, ``The indeterminacy of ecological interactions as perceived through
  perturbation experiments,'' {\em Ecology}, vol.~69, no.~2, pp.~508--515,
  1988.

\bibitem{NA92}
H.~Nakajima, ``Sensitivity and stability of flow networks,'' {\em Ecological
  Modelling}, vol.~62, no.~1-3, pp.~123--133, 1992.

\bibitem{SRTC05}
A.~Saltelli, M.~Ratto, S.~Tarantola, and F.~Campolongo, ``Sensitivity analysis
  for chemical models,'' {\em Chemical Reviews}, vol.~105, no.~7,
  pp.~2811--2828, 2005.

\bibitem{ShiFei09}
G.~Shinar, U.~Alon, and M.~Feinberg, ``Sensitivity and robustness in chemical
  reaction networks,'' {\em SIAM Journal on Applied Mathematics}, vol.~69,
  no.~4, pp.~977--998, 2009.

\bibitem{ShiFei10}
G.~Shinar and M.~Feinberg, ``Structural sources of robustness in biochemical
  reaction networks,'' {\em Science}, vol.~327, no.~5971, pp.~1389--1391, 2010.

\bibitem{ShiFei11}
G.~Shinar and M.~Feinberg, ``Design principles for robust biochemical reaction
  networks: what works, what cannot work, and what might almost work,'' {\em
  Mathematical Biosciences}, vol.~231, no.~1, pp.~39--48, 2011.

\bibitem{PeF20}
B.~Pascual-Escudero and E.~Feliu, ``Local and global robustness in systems of
  polynomial equations,'' {\em \emph{To appear in} Mathematical Methods in the
  Applied Sciences}, 2021.

\bibitem{Capet20}
D.~Cappelletti, A.~Gupta, and M.~Khammash, ``A hidden integral structure endows
  absolute concentration robust systems with resilience to dynamical
  concentration disturbances,'' {\em Journal of the Royal Society Interface},
  vol.~17, no.~171, p.~20200437, 2020.

\bibitem{Shietal11}
G.~Shinar, A.~Mayo, H.~Ji, and M.~Feinberg, ``Constraints on reciprocal flux
  sensitivities in biochemical reaction networks,'' {\em Biophysical Journal},
  vol.~100, no.~6, pp.~1383--1391, 2011.

\bibitem{Sontag14}
E.~D. Sontag, ``A technique for determining the signs of sensitivities of
  steady states in chemical reaction networks,'' {\em IET systems biology},
  vol.~8, no.~6, pp.~251--267, 2014.

\bibitem{Gio16}
G.~Giordano, C.~C. Samaniego, E.~Franco, and F.~Blanchini, ``Computing the
  structural influence matrix for biological systems,'' {\em Journal of
  Mathematical Biology}, vol.~72, no.~7, pp.~1927--1958, 2016.

\bibitem{Feliu19}
E.~Feliu, ``Sign-sensitivities for reaction networks: an algebraic approach.,''
  {\em Mathematical biosciences and engineering: MBE}, vol.~16, no.~6,
  pp.~8195--8213, 2019.

\bibitem{MF15}
A.~Mochizuki and B.~Fiedler, ``Sensitivity of chemical reaction networks: a
  structural approach. 1. {E}xamples and the carbon metabolic network,'' {\em
  Journal of Theoretical Biology}, vol.~367, pp.~189--202, 2015.

\bibitem{FM15}
B.~Fiedler and A.~Mochizuki, ``Sensitivity of chemical reaction networks: a
  structural approach. 2. {R}egular monomolecular systems,'' {\em Mathematical
  Methods in the Applied Sciences}, vol.~38, no.~16, pp.~3519--3537, 2015.

\bibitem{BF18}
B.~Brehm and B.~Fiedler, ``Sensitivity of chemical reaction networks: a
  structural approach. 3. {R}egular multimolecular systems,'' {\em Mathematical
  Methods in the Applied Sciences}, vol.~41, no.~4, pp.~1344--1376, 2018.

\bibitem{VM17}
N.~Vassena and H.~Matano, ``Monomolecular reaction networks: Flux-influenced
  sets and balloons,'' {\em Mathematical Methods in the Applied Sciences},
  vol.~40, no.~18, pp.~7722--7736, 2017.

\bibitem{V17}
N.~Vassena, ``Sensitivity of monomolecular reaction networks: signed
  flux-response to reaction rate perturbations,'' in {\em 2017 European
  Conference on Circuit Theory and Design (ECCTD)}, IEEE, 2017.

\bibitem{V21}
N.~Vassena, ``Sensitivity of metabolic networks: perturbing metabolite
  concentrations,'' {\em arXiv preprint arXiv:2012.10687}, 2020.

\bibitem{Vas20}
N.~Vassena, {\em Sensitivity of Metabolic Networks}.
\newblock PhD thesis, Freie Universit{\"a}t in Berlin, 2020.

\bibitem{VGB20}
N.~Vassena, ``Good and bad children in metabolic networks,'' {\em Mathematical
  Biosciences and Engineering}, vol.~17, no.~6, pp.~7621--7644, 2020.

\bibitem{Fei19}
M.~Feinberg, {\em Foundations of Chemical Reaction Network Theory}.
\newblock Springer, 2019.

\bibitem{HJ13}
R.~A. Horn and C.~R. Johnson, {\em Matrix Analysis}.
\newblock Cambridge University Press, 2013.

\bibitem{Rock69}
R.~T. Rockafellar, ``The elementary vectors of a subspace of
  $\mathbb{R}^{n}$,'' in {\em Proceedings of the Chapel Hill Conf.}, University
  of North Carolina Press, 1969.

\bibitem{SchHil94}
S.~Schuster and C.~Hilgetag, ``On elementary flux modes in biochemical reaction
  systems at steady state,'' {\em Journal of Biological Systems}, vol.~2,
  no.~02, pp.~165--182, 1994.

\bibitem{Kl17}
S.~Klamt, G.~Regensburger, M.~P. Gerstl, C.~Jungreuthmayer, S.~Schuster,
  R.~Mahadevan, J.~Zanghellini, and S.~M{\"u}ller, ``From elementary flux modes
  to elementary flux vectors: Metabolic pathway analysis with arbitrary linear
  flux constraints,'' {\em PLoS Computational Biology}, vol.~13, no.~4,
  p.~e1005409, 2017.

\end{thebibliography}

\end{document}